\documentclass[apj]{emulateapj}
\usepackage{mathptmx}

\def\gtorder{\mathrel{\raise.3ex\hbox{$>$}\mkern-14mu
             \lower0.6ex\hbox{$\sim$}}}
\def\ltorder{\mathrel{\raise.3ex\hbox{$<$}\mkern-14mu
             \lower0.6ex\hbox{$\sim$}}}

\journalinfo{The Astrophysical Journal, 
???: ??1--??7, 2004 Month, astro-ph/0312594}
\slugcomment{Received 2003 Dec 7; Accepted 2003 Jan 16}

\shorttitle{DETAILED OPTICAL LIGHT CURVE OF GRB 030329}

\shortauthors{Y. LIPKIN, E. O. OFEK, A. GAL-YAM, ET AL.}

\begin{document}

\title{The detailed optical light curve of GRB~030329}

\author{
Y.~M.~Lipkin,\altaffilmark{1}$^{,}$\altaffilmark{*}
E.~O.~Ofek,\altaffilmark{1}$^{,}$\altaffilmark{*}
A.~Gal-Yam,\altaffilmark{1}$^{,}$\altaffilmark{*}
E.~M.~Leibowitz,\altaffilmark{1}
D.~Poznanski,\altaffilmark{1}
S.~Kaspi,\altaffilmark{1}
D.~Polishook,\altaffilmark{1}
S.~R.~Kulkarni,\altaffilmark{2}
D.~W.~Fox,\altaffilmark{2}
E.~Berger,\altaffilmark{2}
N.~Mirabal,\altaffilmark{3}
J.~Halpern,\altaffilmark{3}
M.~Bureau,\altaffilmark{3}
K.~Fathi,\altaffilmark{4}$^{,}$\altaffilmark{5} 
P.~A.~Price,\altaffilmark{6}
B.~A.~Peterson,\altaffilmark{6}
A.~Frebel,\altaffilmark{6}
B.~Schmidt,\altaffilmark{6}
J.~A.~Orosz,\altaffilmark{7}
J. B.~Fitzgerald,\altaffilmark{7}
J.~S.~Bloom,\altaffilmark{8}
P.~G.~van~Dokkum,\altaffilmark{9}
C.~D.~Bailyn,\altaffilmark{9}
M.~M.~Buxton,\altaffilmark{9}
and
M.~Barsony\altaffilmark{10}$^{,}$\altaffilmark{11}
}

\altaffiltext{1}{School of Physics and Astronomy and the Wise Observatory, Tel-Aviv University, Tel-Aviv 69978, Israel. Electronic address: (yiftah, eran, avishay)@wise.tau.ac.il}
\altaffiltext{2}{Division of Physics, Mathematics and Astronomy, 105-24, California Institute of Technology, Pasadena, CA 91125, USA}
\altaffiltext{3}{Columbia Astrophysics Laboratory, Columbia University, 550 West 120th Street, New York, NY 10027, USA}
\altaffiltext{4}{School of Physics and Astronomy, University of Nottingham, Nottingham, NG7 2RD, UK}
\altaffiltext{5}{Kapteyn Astronomical Institute, P.O. Box 800, 9700 AV Groningen, The Netherlands}
\altaffiltext{6}{Research School of Astronomy and Astrophysics, The Australian National University, Mt. Stromlo Observatory, Weston Creek, P.O., A.C.T. 2611, Australia}
\altaffiltext{7}{Department of Astronomy, San Diego State University, 5500 Campanile Drive, San Diego, CA 92182, USA}
\altaffiltext{8}{Harvard Society of Fellows, 78 Mount Auburn Street, Cambridge, MA 02138.; Harvard-Smithsonian Center for Astrophysics, MC 20, 60 Garden Street, Cambridge, MA 02138, USA}
\altaffiltext{9}{Department of Astronomy, Yale University, New Haven, CT 06520-8101, USA}
\altaffiltext{10}{Department of Physics and Astronomy, San Francisco State University, 1600 Holloway Avenue, San Francisco, CA 94132-4163, USA}
\altaffiltext{11}{Space Science Institute, 3100 Marine Street, Suite A353, Boulder, CO 80303-1058, USA}
\altaffiltext{*}{The first three authors contributed equally to this work.}

\begin{abstract}

We present densely sampled $BVRI$ light curves of the optical transient
associated with the gamma-ray burst GRB~030329, the result
of a coordinated observing campaign conducted at five observatories.
Augmented with published observations of this GRB,
the compiled optical dataset contains $2687$ photometric measurements,
obtained between 78 minutes and 79 days after the burst.
This dataset allows us to follow the photometric evolution of the
transient with unprecedented detail.
We use the data to constrain the light curve of the underlying
supernova 2003dh, and show that it evolved faster than, and was probably
somewhat fainter than the type Ic SN~1998bw, associated with GRB~980425.
We find that our data can be described by a broken power-law decay perturbed
by a complex variable component. 
The early- and late-time decay slopes are determined to be
$\alpha_1\approx1.1$ and $\alpha_2\approx2$.
Assuming this single-break power-law model, we constrain the break to lie
between $\sim3$ and $\sim8$ days after the burst.
This simple, singly-broken power-law model, derived only from the
analysis of our optical observations, may also account for available
multi-band data, provided that the break happened $\sim8$ days after
the burst.
The more complex double-jet model of Berger et al. provides a
comparable fit to the optical, X-ray, mm and radio observations
of this event.
The unique early coverage available for this event allows us to trace
the color evolution of the afterglow during the
first hours after the burst. We detect
a significant change in optical colors during the first day.
Our color analysis is consistent with a cooling break frequency 
sweeping through the optical band during the first day.
The light curves of GRB 030329 reveal a rich array of variations,
superposed over the mean power-law decay.
We find that the early variations ($\ltorder 8$ days after the burst)
are asymmetric, with a steep rise followed by a relatively slower (by a
factor of about two) decline.
The variations maintain a similar time scale during the
first four days, and then get significantly longer.
The structure of these variations is similar to those previously
detected in the afterglows of several GRBs. 

\end{abstract}


\section{Introduction}
\label{introduction}

Following the discovery of low-energy transients associated with long 
duration ($> 2$ s) gamma-ray bursts (GRBs; van Paradijs et al. 1997),
a major effort was made to characterize the temporal evolution
 of these sources across the electromagnetic spectrum.
In the optical regime, the associated transient sources were
found to decline rapidly with time.
The emission from the first optical transients (OTs) discovered was
reported to decay as a power-law with time, extending from the epoch
of discovery and continuing for tens of days (e.g., GRB 970228 --
Galama et al. 1997; GRB 970508 -- Galama et al. 1998a; Sokolov et
al. 1998; Bloom et al. 1998).

The decay slopes measured for OTs were used to constrain 
explosion models.
In particular, with the increasing popularity of non-isotropic models
(involving highly relativistic jets or cones), a temporal break in the
decline slope was predicted (e.g., Rhoads 1997; Panaitescu,
M\'esz\'aros, \& Rees 1998; Sari, Piran \& Halpern 1999) -- and
detected (e.g., Castro-Tirado et al. 1999; Kulkarni et al. 1999;
Stanek et al. 1999; Harrison et al. 1999; Price et al. 2001).

Following the discovery of supernova (SN) 1998bw in the error box 
of GRB 980425 (Galama et al. 1998b) the association of GRBs with SN 
explosions came into focus.
Late-time ``bumps'' in OT light curves were interpreted as the
signature of underlying SN explosions (e.g., Bloom et al. 1999).
While observational evidence supporting the SN hypothesis accumulated
(e.g., Bloom et al. 2002; Garnavich et al. 2003; Price et. al. 2003a),
a direct observational proof for the existence of an underlying SN
explosion has long remained missing, and alternative explanations for
the origin of these bumps  were suggested (e.g., Waxman  \& Draine
2000; Esin \& Blandford 2000; Reichart 2001).

With growing interest and efforts by the astronomical community,
the observed OT light curves became increasingly better-measured. 
In particular, for some OTs, a dense temporal sampling of the light curve,
sometimes starting shortly (minutes to hours) after the GRB trigger,
was carried out by world-wide observing networks. A case to point is the
OT of GRB 021004, where multiple bumps and wiggles in the light curve,
probably not associated with a SN, were first observed
(e.g.,  Bersier et al. 2003; Mirabal et al. 2003; Fox et al. 2003).    
 
On 2003 March 29, a bright GRB was detected by the HETE-II spacecraft 
(Vanderspek et al. 2003). The early discovery of the associated OT 
(Peterson \& Price 2003; Torii 2003; Torii et al. 2003; Price et
al. 2003b; Uemura 
et al. 2003; Sato et al. 2003), and its brightness, triggered a 
world-wide observational effort, involving tens of ground and space-based
facilities, observing in various wavelengths, from X-ray to radio.
The brightness of the OT allowed the prompt determination of the
redshift of the source.
At z=0.1685 (Greiner et al. 2003a) this event is the closest 
GRB to date for which a typical OT was discovered.
The relatively low redshift of this burst presented a unique
opportunity to search for a clear spectroscopic signature of an
underlying SN.
Indeed, intensive spectroscopic monitoring of the optical source
revealed the emerging spectrum of a type Ic SN 1998bw-like event, designated
SN 2003dh, conclusively
proving that at least some of the long-duration GRBs are associated
with SN explosions (Stanek et al. 2003; Hjorth et al. 2003; Matheson
et al. 2003). 

Numerous works have already presented observations of this unique event.
Early optical observations, obtained with small telescopes shortly
after the burst, were reported by Uemura et al. (2003), Torii et al. (2003), 
Smith et al. (2003), Sato et al. (2003), and Urata et al. (2003). 
Burenin et al. (2003), Bloom et al. (2003) and Matheson
et al. (2003) presented the results of multi-band follow-up campaigns in the
optical and near-IR. 
Greiner et al. (2003b) presented optical polarization monitoring of
GRB 030329, detecting significant variability.
These works mostly relied on data collected at a single
geographical location, thus limiting their ability to achieve a continuous 
temporal coverage. Intensive monitoring of the afterglow of GRB 030329 in the
radio and millimeter wavelengths was reported by Berger et al. (2003) and 
Sheth et al. (2003), while X-ray observations were reported by Tiengo
et al. (2003).

In this paper, we report the results of an intensive, coordinated,
world-wide campaign designed to follow the light curve of the OT
associated with GRB 030329.
Combining data from five observatories in three continents, we
achieved an almost continuous coverage of the OT during the first few
days after the GRB.
Careful cross-calibration was used to bring data collected using many
different instruments to the same reference system.
This internally-consistent data set has allowed us to correctly
incorporate four other sets of observations now available in the
literature.
The final light curves which we compiled are unprecedented in their temporal
sampling, and reveal a uniquely rich and complicated photometric
evolution.
We assume throughout a cosmology with $\Omega_{m}=0.3$, $\Omega_{\Lambda}=0.7$, $H_{0}=65$~km~s$^{-1}$~Mpc$^{-1}$.

\section{Observations and Data Reduction}
\label{reduction}
We performed time-resolved CCD photometry of the OT at five 
observatories, from March $29.72$ to June $17.19$ UT ($5.66$~hr, to $79.7$~d
after the burst).
Observations were carried out through $B$, $V$, $R$, and
$I$ filters, accumulating a total of 77, 104, 928, and 96 data
points, respectively.
Details of the equipment and a summary of the observations are given
in Table~\ref{TableObservations}.

\begin{deluxetable}{llll}
\tablecolumns{4}
\tablewidth{0pt}
\tablecaption{List of observatories\label{TableObservations}}
\tablehead{
\colhead{ID} &
\colhead{Observatory+Instrument} &
\colhead{E. long.} &
\colhead{Filter (number of observations)}
}
\startdata
 $01$  & Wise~$1$m+Tek              &   $34.8$ & B(9), V(25), R(484), I(25)   \\
 $02$  & Wise~$1$m+SITe             &   $34.8$ & B(8),  V(7),  R(57)          \\
 $03$  & SSO~$1$m+WFI               &  $149.1$ & B(5), V(2), R(5), I(2)       \\
 $05$  & MDM~$1.3$m+$2.4$m+SITe     & $-111.6$ & B(1), V(1), R(311), I(1)     \\
 $06$  & Palomar~$1.5$m+Norris      & $-116.9$ & I(3)                         \\
 $07$  & Mt. Laguna~$1$m+Loral        & $-116.4$ & B(54), V(69), R(71), I(67)   \\
\hline\hline
$21$  & Kyoto\tablenotemark{a}      & $\sim135$& CR(391)\tablenotemark{e}\\
$31$  & RTT\tablenotemark{b}        &  $30.3$  & B(144), V(167), R(168), 
I(165)\\
$41$  & CTIO 1.3\tablenotemark{c}   &  $-70.8$ & B(9), V(13), I(13)            
\\
$51$  & FLWO\tablenotemark{d}       & $-110.9$ & B(62), V(57), R(111), I(57)\\
$52$  & KAIT\tablenotemark{d}       & $-121.6$ & B(14), V(15), R(15), I(15)\\
$53$  & LCO\tablenotemark{d}        &  $-70.7$ & B(4), V(4), I(4)\\
$54$  & LCO-40\tablenotemark{d}     &  $-70.7$ & R(2)\\
$55$  & KPNO4m\tablenotemark{d}     & $-111.6$ & B(19), R(4)\\
$56$  & Magellan2\tablenotemark{d}  &  $-70.7$ & R(25)\\
$57$  & Dupont\tablenotemark{d}     &  $-70.7$ & B(4)\\
 \enddata
\tablecomments{The upper part of the table lists the observatories which took part in our 
   campaign. The lower part lists external data sources.}
\tablenotetext{a}{Several observatories in Japan. Uemura et al. (2003).}
\tablenotetext{b}{Burenin et al. (2003).}
\tablenotetext{c}{Bloom et al. (2003).}
\tablenotetext{d}{Matheson et al. (2003).}
\tablenotetext{e}{Unfiltered observations transformed to $R$-band, see Uemura et al. (2003).}
\end{deluxetable}

The images were bias-subtracted and flat-field corrected in the
standard fashion.
In each frame we measured the magnitude of the OT, as well as of several
reference stars, with an aperture of $2$~arcsec radius. 
Three to nine reference stars were measured in each subset,
depending on the depth and the field of view of the images.
All of the reference stars were tested to be non-variable.
The reference stars used in each subset are listed in
Table~\ref{TableReferenceStars}. 

\begin{deluxetable}{cllllllll}
\tablecolumns{9}
\tablewidth{0pt}
\tablecaption{Reference Stars \label{TableReferenceStars}}
\tablehead{
\colhead{Coordinates (2000)} &
\colhead{Number\tablenotemark{a}} &
\colhead{$V$ mag\tablenotemark{a}} &
\multicolumn{6}{c}{Used in sets\tablenotemark{b}} 
}
\startdata
10:44:28.62  +21:27:45.4 & 005 & 15.587 & &2&3& &6& \\
10:44:36.86  +21:26:59.1 & 016 & 13.266 & & & & &6& \\
10:44:39.07  +21:30:59.1 & 019 & 17.616 &1& &3& &6&7\\
10:44:39.86  +21:34:15.0 & 021 & 16.842 &1& &3& &6& \\
10:44:41.75  +21:31:52.6 & 026 & 19.331 & & &3&5& &7\\
10:44:42.02  +21:32:32.1 & 027 & 16.839 & &2&3&5& &7\\
10:44:48.03  +21:34:18.5 & 037 & 17.909 &1&2&3& & & \\
10:44:53.66  +21:30:12.1 & 047 & 18.396 & & & & &6& \\
10:44:54.45  +21:34:29.2 & 049 & 14.136 & & &3& & & \\
10:44:54.99  +21:29:46.3 &     & 20.275\tablenotemark{c} & & & &5& & \\
10:44:55.00  +21:31:42.9 & 050 & 19.598 & & & & & &7\\
10:45:06.48  +21:36:13.7 & 093 & 16.245 & &2& & & & \\
10:45:09.81  +21:35:10.2 & 098 & 15.469 & &2&3& & & \\
10:45:15.36  +21:34:16.1 & 110 & 14.730 & &2&3& &6& \\

 \enddata
\tablenotetext{a}{Star numbers and magnitudes refer to the photometry
  of Henden (2003).} 
\tablenotetext{b}{Observatory IDs given in
  Table~\ref{TableObservations}.} 
\tablenotetext{c}{Photometry measured in our dataset, and scaled to
  the system of Henden (2003)}
\end{deluxetable}

For each of the detectors used, we obtained a set of
internally-consistent magnitudes of the OT by minimizing the scatter
of the reference stars over the subset.
Outlying measurements were removed during the process.
Apparently underestimated errors in the photometry of
reference stars were increased so that $\chi^{2}/dof=1$
(where $dof$ is the number of degrees of freedom)
for each of the reference stars in the subset.

Cross calibration, including color terms,
of the different instruments was carried out
by transforming the photometric system of each of the
subsets to the one of Henden (2003).
The transformation parameters were derived by linearly fitting 
the differences between the weighted-mean
magnitude of the reference stars, and their magnitudes in the  Henden
(2003) system, to the colors of the reference stars reported by this author.
Because of the initial uncertainty in the OT colors, two iterations were
required to transform its magnitude to the reference system.
The transformation uncertainty of each data subset was added in quadrature
to the measured photometric errors.
The photometric errors do not include the uncertainty in the
zero point of the calibration, $0.02$~mag (Henden 2003).

The photometric measurements obtained at the Mount Laguna Observatory failed
to yield consistent color terms and were therefore cross-calibrated by
eliminating the offset between overlapping segments of this subset and
the rest of the light curve. From the residual scatter in
these overlapping segments, we estimate  
the resultant systematic cross-calibration errors of this set to be $\sim5\%$.

We augmented our data set with external photometry from 
Burenin et al. (2003), Uemura et al. (2003), and Bloom et al. (2003),
kindly provided in digital form by the authors, as well as
with the data published by Matheson et al. (2003).
Each of the external data sources was cross-calibrated with our set
using the procedure described above for the Mount Laguna Observatory
data. 
Since the color of the OT changed with time, this process may have
introduced small systematic errors.
From the scatter between the external sets and our light curves, we
estimate this error to be smaller than $5\%$. 

A few data segments from Uemura et al. (2003), temporally overlapping
with our own, were not incorporated into
our light curve because the photometric errors in these segments (and
hence the scatter of the data points) were greater than  in our data.
We note, however, that our data and those of Uemura et al. are in good
agreement throughout the overlapping segments.
The $I$-band light curve of Burenin et al. (2003), covering the time-span
between $0.2$~d and $0.6$~d, does not overlap with our $I$-band
data.
The zero point for this segment was set by coarsely extrapolating later
$I$-band segments, and should therefore be regarded with caution.

The earliest available observations of the OT ($\sim1.3$~hr after the burst)
were obtained by two groups. A set
of unfiltered observations were reported by Uemura et al. (2003), who
transformed the resulting magnitudes to the standard $R$-band.
After cross calibration, these data are in good agreement with the rest of our data. 
A couple of measurements in standard $R$ were taken through clouds with the SSO
1m telescope (Price et al. 2003b).
After cross-calibration, the two points yielded significantly brighter
measurements, by about $0.1$~mag.
Other observations from the SSO 1m telescope, taken under
better conditions, are in good agreement with the rest of the data.
Because of the good agreement between the bulk of our data and those of
Uemura et al. (2003),
and because of the unfavorable weather conditions under which the
early SSO images
were obtained, we preferred the light curve of the former over the latter
two points.
The results we present below, however, are insensitive to this
selection, and would remain essentially the same had we corrected
Uemura's data to fit the SSO points.
The recently reported early observations by Torii et al. (2003)
are even brighter than the SSO points. Thus, it appears that 
forming a consistent picture of the early light curve of this event will
require further analysis, once all the relevant data are available. 

Altogether, we compiled 2687 photometric measurements (333, 360, 1644,
and 350 in $B$, $V$, $R$, and $I$, respectively), making GRB 030329
the most extensively studied extragalactic cosmic explosion in the
optical regime, with the possible exception of SN 1987A.
The complete set of photometric measurements presented in this paper are 
listed in
Table~\ref{TableLC}\footnote{The complete Table with all $2687$ measurements is 
given in
  the electronic version of this paper. The data are also available from
  our web site: http://wise-obs.tau.ac.il/GRB030329/}.

\begin{deluxetable}{lllllll}
\tablecolumns{7}
\tablewidth{0pt}
\tablecaption{Observations \label{TableLC}}
\tablehead{
\colhead{(1)} &
\colhead{(2)} &
\colhead{(3)} &
\colhead{(4)} &
\colhead{(5)} &
\colhead{(6)} &
\colhead{(7)}
}
\startdata
01      & 04    & 2729.21443    & 1.23024       & 16.438        & 0.018 & 
0.050\\
01      & 04    & 2729.21597    & 1.23177       & 16.414        & 0.020 & 
0.051\\
01      & 04    & 2729.21749    & 1.23329       & 16.447        & 0.018 & 
0.050\\
01      & 04    & 2729.21900    & 1.23481       & 16.398        & 0.019 & 
0.050\\
01      & 04    & 2729.22052    & 1.23632       & 16.415        & 0.018 & 
0.050\\
01      & 04    & 2729.22203    & 1.23783       & 16.432        & 0.018 & 
0.050\\
01      & 04    & 2729.22355    & 1.23936       & 16.434        & 0.018 & 
0.050\\
01      & 03    & 2729.22956    & 1.24536       & 16.786        & 0.012 & 
0.022\\
01      & 05    & 2729.23286    & 1.24866       & 15.940        & 0.011 & 
0.060\\
01      & 04    & 2729.23549    & 1.25129       & 16.416        & 0.013 & 
0.048\\
\enddata  
\tablecomments{The first ten data points of our light curve. The columns
 are:
(1) the observatory code (see Table~\ref{TableObservations}); 
(2) the filter used. $2$,$3$,$4$,$5$ are for $B$, $V$, $R$, $I$, respectively;
(3) $JD-2450000$;
(4) time since $t_{burst}=2452727.98419757$;
(5) calibrated magnitude;
(6) the error in magnitude including the self calibration error of the
individual observatory;
(7) the error in magnitude including the uncertainty in cross
calibrating the various observatories.
}
\end{deluxetable}

\section{The light curve}
\label{SecLC}
\begin{figure*}
\centerline{\includegraphics[width=18cm]{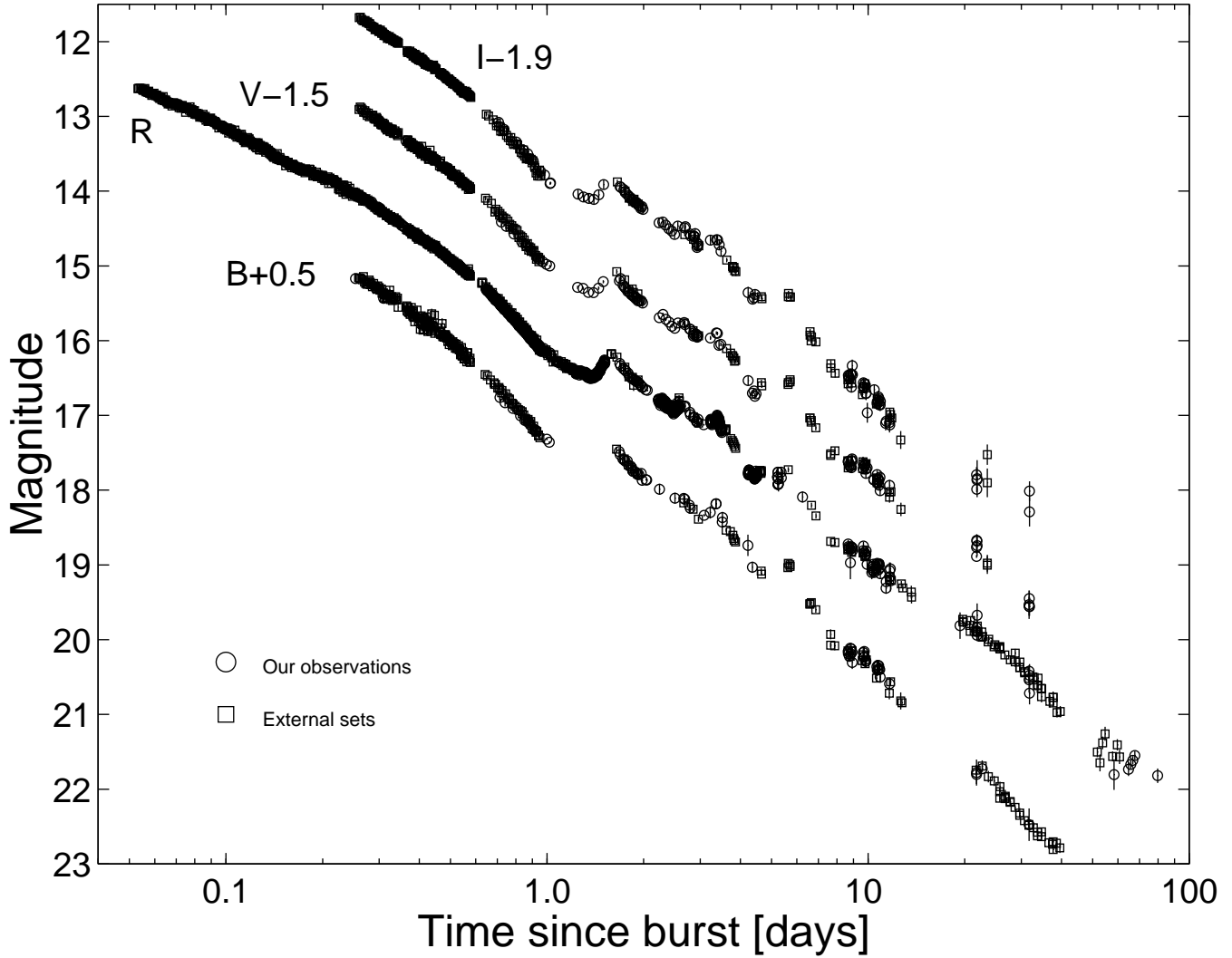}}
\caption{$BVRI$ light curve of GRB~030329. Our observations are marked
 by circles, and the external data sets (see
  Table~\ref{TableObservations}), cross-calibrated to our photometric
  system, are marked by squares.
  For presentation purposes the $BVI$ light curves are shifted vertically by
  $+0.5$, $-1.5$, and $-1.9$~mag, respectively.
}
\label{FigLC}
\end{figure*}

The final $BVRI$ light curves of GRB 030329 are presented in Figure~\ref{FigLC}.
Our best-sampled set is the $R$-band light curve, with an
almost continuous coverage during $0.05 \ltorder t \ltorder 3$~d,
and a relatively dense sampling up to $t\approx67$~d.
We therefore focus our attention on this light curve, and,
augment our analysis with results from other bands when necessary.

While the OT grossly follows a power-law declining trend,
it shows remarkably strong short-term deviations from the smooth,
monotonic decline, particularly between $0.05 \ltorder t \ltorder
5$~d.
Later periods, between $8 \ltorder t \ltorder 14$~d, and at
$t\gtorder50$, also feature pronounced variations of the light curve.
We note that the light curves do not show
a priori any clear distinction between ``early'' and ``late'' power law
components -- the signature of a break in the decline rate.
We also note the lack of a clear bump which may be attributed
to SN 2003dh, although its significant contribution to the total
brightness of the OT was established spectroscopically.

At $R=23.25\pm0.15$~mag (Fruchter 2003, Priv. Comm.), the host becomes a
significant contributor to the observed flux during the late decline 
phase of the OT.
Hence below, unless stated otherwise, we shall discuss the R-band
light curve of the OT only, obtained by subtracting the host flux from
our measurements.
The photometric errors of the OT were modified to include the
uncertainty in the host galaxy magnitude.

To further discuss the light curve, we decompose it into a SN
component, and an ``afterglow'' component, which is further divided into two
components: a smooth, monotonic  decline component (the main observational
characteristic of GRB afterglows), and a perturbations component --
variations over different time scales and intensities about the
smooth decay of GRB 030329.
In the following sections we shall consider each of the OT
components separately.

\subsection{The supernova component}
\label{SecSNLC}

Spectroscopic analysis of SN 2003dh (Stanek et al. 2003; 
Hjorth et al. 2003; Matheson et al. 2003; Chornock et al. 2003) 
revealed a remarkable similarity to the well studied SN 1998bw associated
with GRB 980425 (although, see Mazzali et al. 2003; Kawabata et al. 2003). 
In light of this similarity, we base our investigation
of the SN component in the afterglow of GRB 030329 on the known
properties of SN 1998bw.

We constrain the SN component by comparing
the OT light curve with four alternative models for SN 2003dh. 
All four models are modifications of the light curve of SN 1998bw,
corrected to the redshift of SN 2003dh (z=0.1685).
The transformation was carried out by applying synthetic photometry
(using the methods presented in Poznanski et al. 2002) to the large
collection of SN 1998bw spectra reported by Patat et al. (2001), and 
taking into account the greater luminosity distance of SN
2003dh (810~Mpc, compared to 37~Mpc of SN 1998bw), cosmological
redshift and time-dilation effects, as well as the shift in the SN
spectrum sampled by each filter (K-corrections).
The SN light curve was also corrected for Galactic extinction in the
direction of SN 2003dh ($E_{B-V}=0.025$~mag, Schlegel, Finkbeiner, \&
Davis 1998),
but was not corrected for the host galaxy extinction (see below).
Our derived light curves are in good agreement with those computed by
Bloom et al. (2003) using a slightly different approach.

Comparing the models to the data, we note that any
variation in the light curve should probably be attributed to the afterglow,
since no such variations have been detected in the light curves of any of the well
observed SNe~Ic (e.g., SN 1998bw -- Galama et al. 1998b; McKenzie \&
Schaefer 1999; Patat et al. 2001; SN 2002ap -- Gal-Yam, Ofek \&
Shemmer 2002; Panday et al. 2003; Yoshii et al. 2003; Foley et
al. 2003; SN 1999ex -- Stritzinger et al. 2002; SN 1994I -- Yokoo et
al. 1994).
A valid SN model is therefore required to be fainter than the
bottom of the dips in the light curve.
In particular, the minima of two rebrightening episodes in the $R$-band, one 
of
$\sim0.3$~mag around $t\approx52$~d, and another of $\sim0.2$~mag around
$t\approx68$~d (see \S\ref{SecVarLC}) set upper limits on the SN magnitude
of $R=21.93\pm0.15$ and $R=22.04\pm0.13$ mag at the times of these episodes, respectively.

\begin{figure}
\centerline{\includegraphics[width=9cm]{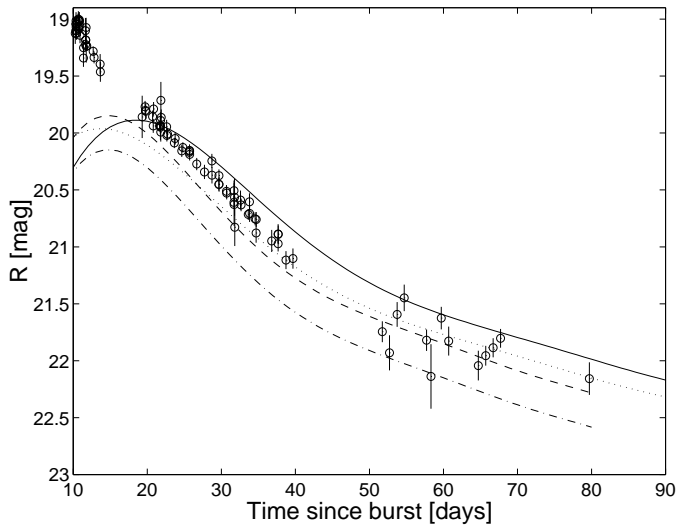}}
\caption{The host-subtracted $R$-band light curve of the OT (open
  circles) compared to four  model light curves of SN 2003dh.
  The plotted lines are: the light curve of SN~1998bw, 
  redshifted and K-corrected
  (solid line);
  the same model, after applying the best fitting magnitude shift and 
  temporal shift relative to the GRB (see text; dotted line); the same model, 
  after applying
  the best fitting magnitude-shift and stretch correction
  (see text; dashed line); and the time-stretched and
  magnitude-shifted model lowered by $0.3$~mag  (dash-dotted line).
}
\label{SN_LC}
\end{figure}

Figure~\ref{SN_LC} shows the $R$-band light curve of the OT (open circles).
Overlaid are the four different model light curves for the supernova component.
The solid line in Fig.~\ref{SN_LC} is the redshift-corrected light
curve of SN 1998bw described above, assuming it exploded simultaneously
with the GRB.
This simple model is consistently brighter than the data points after
day $\sim20$, and is therefore ruled out --
had SN 2003dh been identical to SN 1998bw, the OT would have
been brighter than observed after day $20$.

Hjorth et al. (2003) decomposed their observed
spectra into a SN Ic component (using redshifted versions of SN 1998bw
spectra) and a power-law component, typical for the optical emission
from cosmological GRBs.
Using this method, these authors derived a $V$-band light curve of the SN
component during the first month after the GRB. 
They found that their data could not be fitted by a redshifted and
K-corrected $V$-band light curve of SN 1998bw (see their Fig.~3),
but required that the SN component would be slightly brighter (by
$\sim 0.2$ mag) at the peak, and would also decline much faster than
SN 1998bw, becoming at least $0.7$ mag fainter 28 days after the
GRB. Similar results were obtained by Mazzali et al. (2003), who 
combined spectral analysis with explosion models.

With this in mind, we consider alternative models for the SN component.
Following Hjorth et al. (2003) and motivated by theoretical models of delayed
core-collapse (e.g., Vietri \& Stella 1998; Berezhiani et al. 2003), we
consider a non-simultaneous SN model.
The model light curve (dotted line) was obtained by fitting the $V$-band light curve of
SN 1998bw to that of SN 2003dh (Hjorth et al. 2003), with two free
parameters: a time lag between the GRB and the SN explosion
($\Delta{T}$), and a magnitude-correction ($\Delta{m}$).
The best fit values are: $\Delta{T}=-4.7_{-2.2}^{+1.7}$~d and
$\Delta{m}=0.08\pm0.10$~mag, with $\chi^{2}/dof=2.5/4$.
This model initially agrees with the data, but becomes too bright
after day 50, and is therefore discarded.
Larger temporal offsets (required to make the SN fainter at late
times) are inconsistent with the spectral analysis of Stanek et
al. (2003), Matheson et al. (2003), and Hjorth et al. (2003). 
In the next model (the dashed line in Fig.~\ref{SN_LC}), we
introduced a magnitude correction parameter ($\Delta{m}$), and a
"stretch" parameter $s$ (similar to the formalism used by Perlmutter
et al. 1999) which adjusts the width of the light curve so that the
model luminosity of SN 2003dh at time $t$ after the burst is given by the
luminosity of the redshifted SN 1998bw at $t/s$.
We find $s=0.80\pm0.05$ and $\Delta{m}=-0.01\pm0.10$
(with $\chi^{2}/dof=0.9/4$) by fitting the $V$-band light curve of SN 1998bw to
the measurements of Hjorth et al. (2003).
This model provides a somewhat better fit to the data compared to the
delayed-GRB model (dotted line), but still overestimates the data
after day $50$. 

In search for a SN model which will satisfy all available 
constraints, we consider two additional inputs. 
Bloom et al. (2003) used a self-consistent multicolor data set
obtained with a single instrument (ANDICAM, mounted on the CTIO 1.3m
telescope). 
They found that their data could be adequately modeled using the redshifted and
K-corrected light curves of a SN~1998bw-like event,  which is slightly
brighter than SN 1998bw.
Such light curves are ruled out by our measurements, obtained from day
20 onward.
The probable reason for SN 1998bw being consistent with the
observations of Bloom et al. (2003), is that their data are well sampled
until day 12, and their latest data points were obtained at day 23,
just when our data start indicating that SN 1998bw-like light curves are too
bright.
   
Matheson et al. (2003) did not pursue the full calculation of
K-corrections required to produce the light curves of SN 1998bw as
they would appear for an event exploding at $z=0.1685$.
Instead, they used the $V$-band light curve of SN 1998bw as a proxy for
the $R$-band light curve of SN 2003dh.
Comparing their adopted ``$R$-band'' light curve with the results of our own
calculation, we find that their light curve has a similar shape, but is 
fainter by $\sim0.4$ mag, taking into account their estimated extinction
$A_R<0.2$ mag.
In their SN light curve model, Matheson et al. (2003) further scaled down
SN 2003dh by $\sim0.2$ mag, relative to their adopted SN
1998bw light curves. 
Thus, the good fit they reported for SN 1998bw-like light curves
is actually for an event fainter by $\sim0.6$ mag.

Because our late-time data rule out non-stretched SN 1998bw-like light curves, even if they
are as faint as advocated by Matheson et al. (2003), we are led to construct
a model, which is identical to the stretched model (dashed-line), but
is attenuated by $0.3$~mag (the dash-dotted in Fig.~\ref{SN_LC})
-- a compromise between
the results of Bloom et al. (2003) and Hjorth et al. (2003),
suggesting peak magnitudes at least as bright as those
of SN 1998bw, and those of Matheson et al. (2003), who found
that SN 2003dh was fainter than SN 1998bw by $\sim 0.6$ mag. 
This last model seems to best fit both our
data and other available studies, and is also in good agreement 
with the theoretical modeling reported by Mazzali et al. (2003).
Both the attenuation factor and the stretch parameter of this last
model are well situated in the parameter space of fitted 1998bw-like
SN, which was obtained by Zeh, Klose \& Hartman (2003) for a sample of
seven GRB optical counterparts.
Note, that we do not correct for the host-galaxy extinction
of SN~2003dh. If we assume negligible extinction of SN 1998bw
and the maximal extinction values allowed by the analysis of 
Matheson et al. (2003) and Bloom et al. (2003), $A_{V}\ltorder0.4$,
our adopted peak magnitude may be consistent with that of a SN which is
intrinsically identical to SN~1998bw.
The $\Delta{m}=0.3$~magnitude we adopted for the attenuation of SN2003dh
relative to SN1998bw, is in fact a lower limit on the attenuation
(or conversely, an upper limit on the allowed peak luminosity).
Therefore, below we further check the robustness of
our results to the value of attenuation we adopt in our SN model.

\subsection{The afterglow component}
\label{SecAGLC}

\begin{figure}
\centerline{\includegraphics[width=9cm]{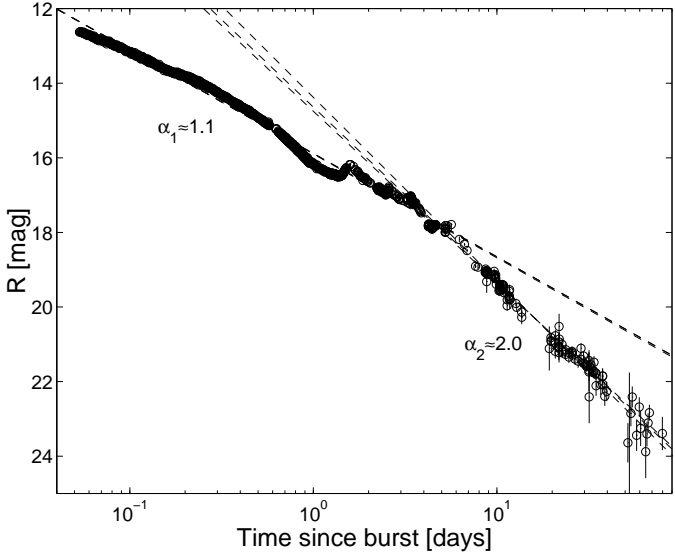}}
\caption{$R$-band light curve of the afterglow. The light curve was derived by
  subtracting the contribution of SN 2003dh and of the host galaxy
  from the observed light curve. Overlaid are a few power-law fits to early
  and late times (see text for details).
A break at $t\sim5$~days is apparent. The slope
  of the pre-break power-law is $\alpha_1\approx1.1$, and that of the
  post-break is $\alpha_2\approx2.0$.
}
\label{FigLCBreak} 
\end{figure}

The clearcut connection between GRB 030329 and SN 2003dh makes it
important to characterize the afterglow in detail, in order to 
check whether this is a typical burst, or an exceptional one.
Figure~\ref{FigLCBreak} shows the $R$-band light curve of the afterglow, 
derived by subtracting the host galaxy flux and the preferred SN model
(dash-dotted line in Figure~\ref{SN_LC}, see \S\ref{SecSNLC} for details).
The afterglow light curve is clearly not a smooth decline - strong variations
are apparent, starting very early after the burst, and continuing throughout
our observations (\S\ref{SecVarLC}).
Inspection by eye shows that the early light curve may be described by a power-law
decline, with a steepening of the slope around $t\sim5$~d. 

We therefore begin our examination by investigating the simplest plausible 
model -- a singly-broken power law. 
We fitted sets of double power laws to the afterglow light curve.
In each fit, we assumed a different break time $t_{break}^{assumed}$
(i.e., one power law was fitted to all the points with $t<t_{break}^{assumed}$
and another one to points with $t>t_{break}^{assumed}$).
The intersection time ($t_\times$) of the two power-laws, and the
sum of the $\chi^2$ values of the two fits were calculated for each 
$t_{break}^{assumed}$.
During the first three days, the light curve was sampled more
intensively than at later times.
To reduce the overwhelming statistical weight of this segment, we
diluted it by binning the points.
The dashed lines in Fig.~\ref{FigLCBreak} are a few examples of such
fits, with $t_{break}^{assumed}=3,5,8$~d.
As can be seen, the values of the early- and late-time slopes
are weakly dependent on the assumed break time. We therefore consider
these values ($\alpha_1\approx1.1$ and $\alpha_2\approx2.0$, where
$\alpha$ is the power-law decay index defined by the dependence of
flux, $f$, on time $f\propto t^{-\alpha}$) to be robustly constrained by the
data.

For assumed break times between $1.5$ and $11$~d, the calculated
intersection between the two power-law fits fall consistently between
$3$ and $5$~d.
Minimum $\chi^2$ values were obtained for assumed breaks between $3$
to $6$~d.
When carried out with light curves in the $B$, $V$ and $I$ bands, the same procedure yielded
a similar range for the intersection points:
$3 \ltorder t_{\times} \ltorder 8$~d. 

To test the sensitivity of our results to the SN model which was
subtracted from the light curve, we repeated the test using several different SN
models.
All the models were redshift- and K-corrected light curves of SN 1998bw, stretched by
$s=0.80$. We varied the values of the attenuation $\Delta{m}$ between 
$0.3\le \Delta{m} \le 0.6$~mag, since brighter values 
($\Delta{m}<0.3$)
are ruled out by our light curves (see \S~\ref{SecSNLC}), and fainter
values ($\Delta{m}>0.6$) are inconsistent with the strength of the
observed SN features in the OT spectra (Matheson et al. 2003; Hjorth
et al. 2003).
The early-time slope did not change
significantly, while the late-time slope  varied between $1.8-2.0$,
leaving the range of intersection times unchanged.

To conclude, for a singly broken power law model,
our data robustly constrain the early- and late-time
decay slopes to lie around $\alpha_1\approx1.1$ and $\alpha_2\approx2.0$, respectively.
These values depend very weakly on the light curve segments used,
or on the SN model subtracted. If we take the intersection time
of these slopes as an estimate for the break time, the resulting values lie around $t\sim5$~d.
However, because of the strong variations in the light curve, 
the time of the break is not well constrained, and could be
placed at any time between $\sim3$ to $\sim8$~d.
If we adopt $t_{break}=5$~d for the time of the break, a fit to the
data before and after the break yields $\alpha_{1}=1.11\pm0.01$
and $\alpha_{2}=1.96\pm0.02$. 
Maintaining self-consistency, the intersection of these two
power laws is $t_{\times}=4.9$~d. 
This result agrees with the late-time power-law slope, $\alpha=2.05$,
measured by Matheson et al. (2003)
from data obtained between $t\approx5$~d to $t\approx61$~d.
It appears that a double-power law, broken somewhere between day
$\sim 3$ and $\sim8$, and strongly 
perturbed by a series of bumps (Figures~\ref{FigLCBreak} and \ref{FigRes07}) 
provides a fair
description of the complex light curve of this OT. We therefore 
continue and analyze the properties of the variable component
derived from the simple broken power-law model in the next section.

Numerous works, analyzing short or sparsely-sampled optical light curves,
have interpreted some of the deviations from a smooth decline observed
in the OT light curves as manifestations of specific phenomena predicted
by popular relativistic jet models. In particular, Berger
et al. (2003) propose a model that attempts to consistently account for 
observations in the radio, mm, optical and X-ray bands. We review
the complex structure seen in our superior data in the context of previous
analyses in section~\S\ref{secbreaks}
below, and confront an updated version of the
Berger et al. (2003) model with our optical observations in
\S\ref{Berger}.
 
\subsection{The variable component}
\label{SecVarLC}
\begin{figure*}
\centerline{\includegraphics[width=18.0cm]{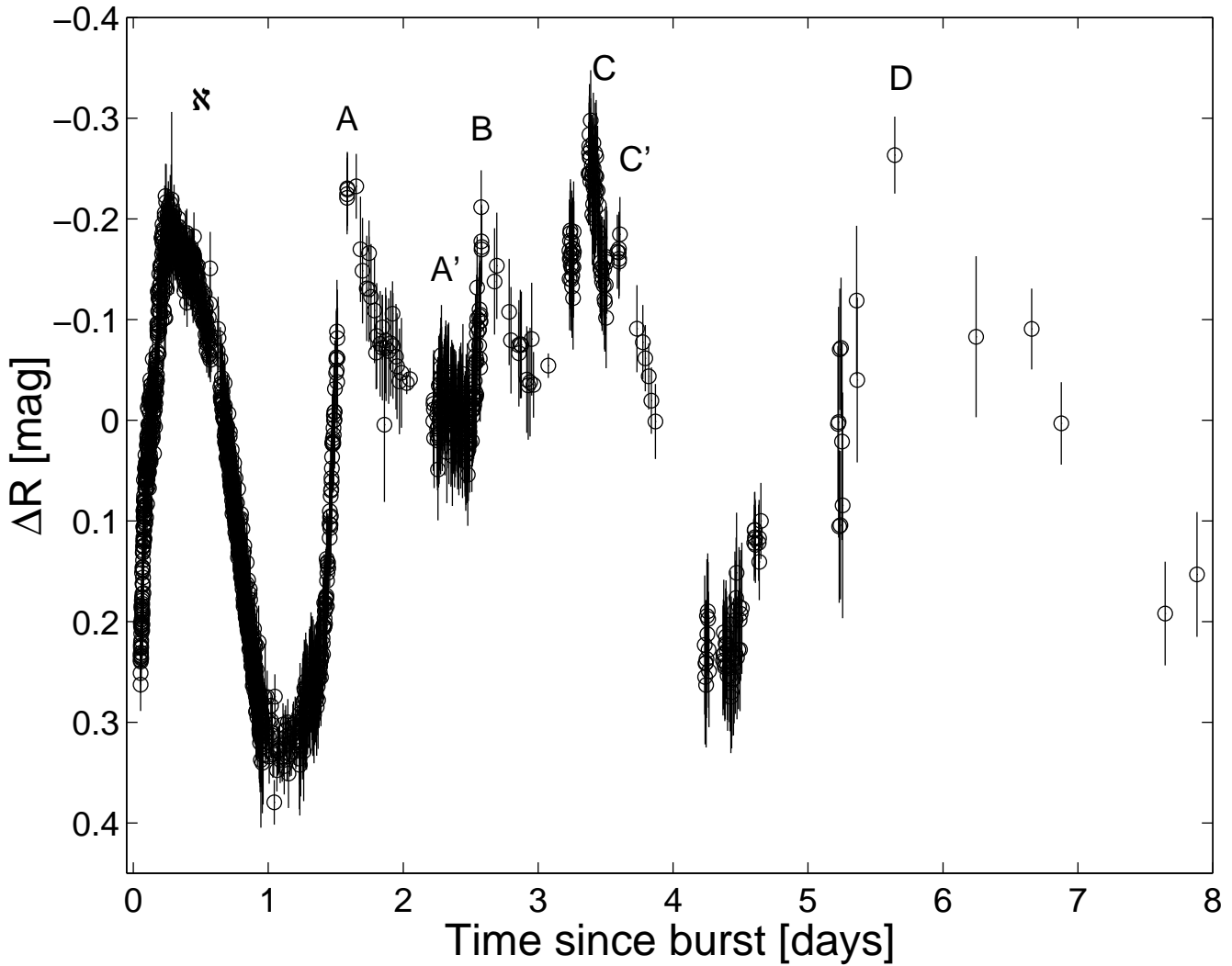}}
\caption{The residual light curve, obtained by subtracting our best
  fit double power-law (see text) from the light curve of the
  afterglow, during the first eight days after the burst. 
  Five strong bumps, along with two possible minor ones, are apparent
  in the light curve.
}
\label{FigRes07}
\end{figure*}

Subtracting our best double power-law decline model 
(\S\ref{SecAGLC}) from the afterglow light curve,
we derive the residual $R$-band light curve of the afterglow. 
Three segments of the residual light curve are shown in
Figures~\ref{FigRes07}, \ref{FigRes812} and \ref{FigRes1842}.

Figure~\ref{FigRes07} shows the residual light curve during the first
eight days after the burst.
Arbitrarily, we consider the variability as a series of bumps, 
although with no model at hand, this is a mere matter of convenience.
The bumps in Fig.~\ref{FigRes07} are marked according to the notation
introduced by Granot, Nakar, \& Piran (2003), who identified four bumps 
in the light curve of the OT,
compiled from reports in GCN circulars ($A$, $B$, $C$, and
$D$ in Fig.~\ref{FigRes07}). The early bump, $\aleph$, was
identified by Uemura et al. (2003), and the minor bumps,  $A'$ and
$C'$ are introduced in this work.
The validity of the minor bump $C'$ is uncertain, because the peak of
the bump is only $\sim1\sigma$ brighter than the dip prior to the
bump, and since the points forming its peak all come from the
same data set (Matheson et al 2003).

The bumps before day $\sim8$, which possibly occurred prior to
the change in the afterglow decline rate (Fig.~\ref{FigRes07}, see
\S~\ref{onebreak} below) are asymmetric in shape,
with an incline that is typically $\sim2$ times as steep
as the decline (with the possible exception of $C'$, which is poorly
sampled).

\begin{figure*}
\centerline{\includegraphics[width=18cm]{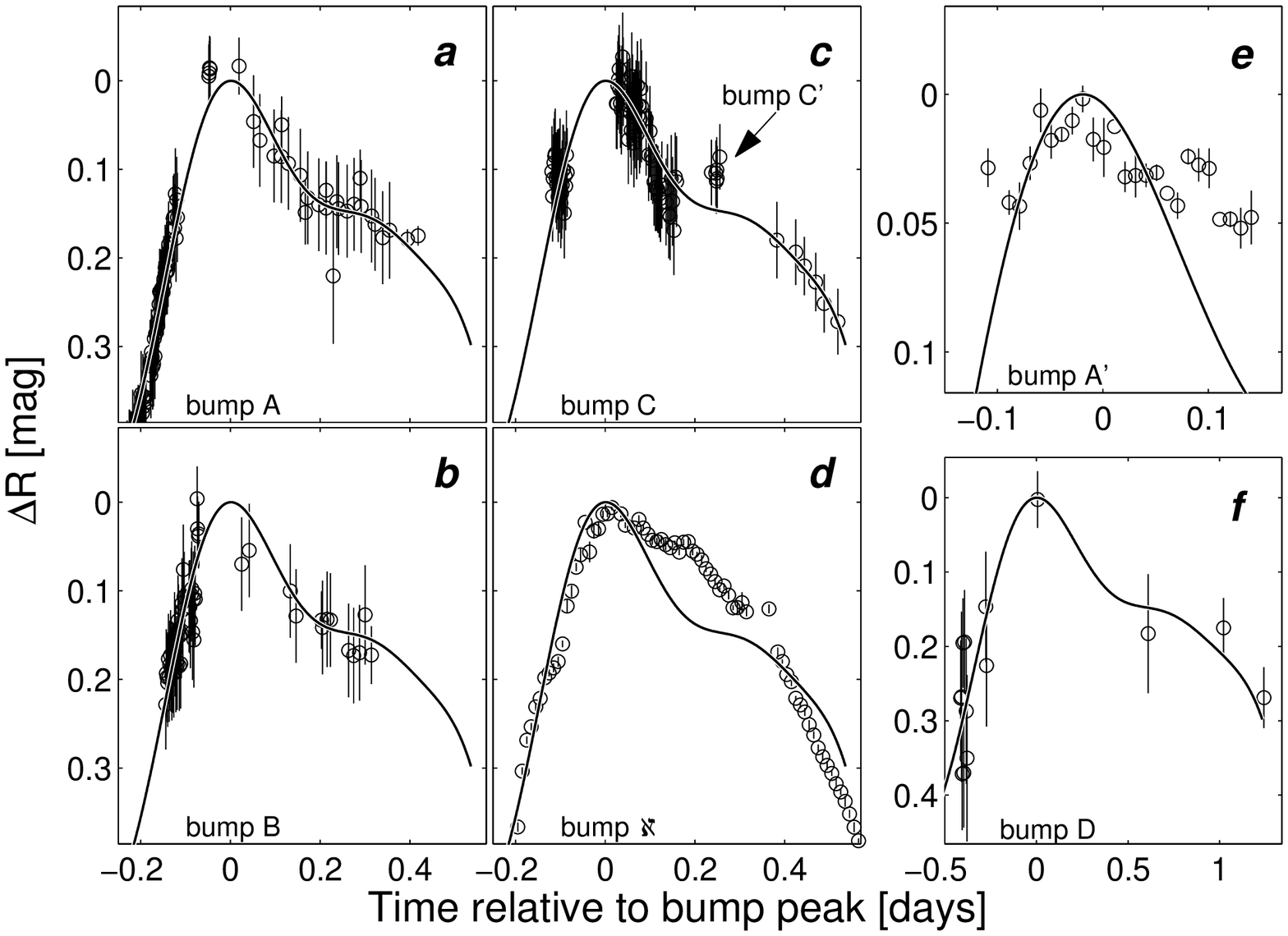}}
\caption{A close up view of the main bumps detected in the early
  residual light curve, which was obtained by subtracting our best
  fit double power-law (see text) from the light curve of the
  afterglow.Panels {\it a}, {\it b}, {\it c}, {\it d}, {\it e}, and
  {\it f}, display bumps $A$, $B$, $C$, $\aleph$, $A'$, and $D$,
  respectively. The light curves in panels {\it d} and {\it e} were 
  binned, to allow convenient browsing. The solid curves are a
  smoothed fit to the superimposed light curves of bumps $A$, $B$, and
  $C$, representing the common structure of these three bumps (see
  text for details). The time scale of the curve in panel~{\it f} was
  stretched by a factor of $2.3$.
}
\label{FigBumps}
\end{figure*}

\begin{figure}
\centerline{\includegraphics[width=8.5cm]{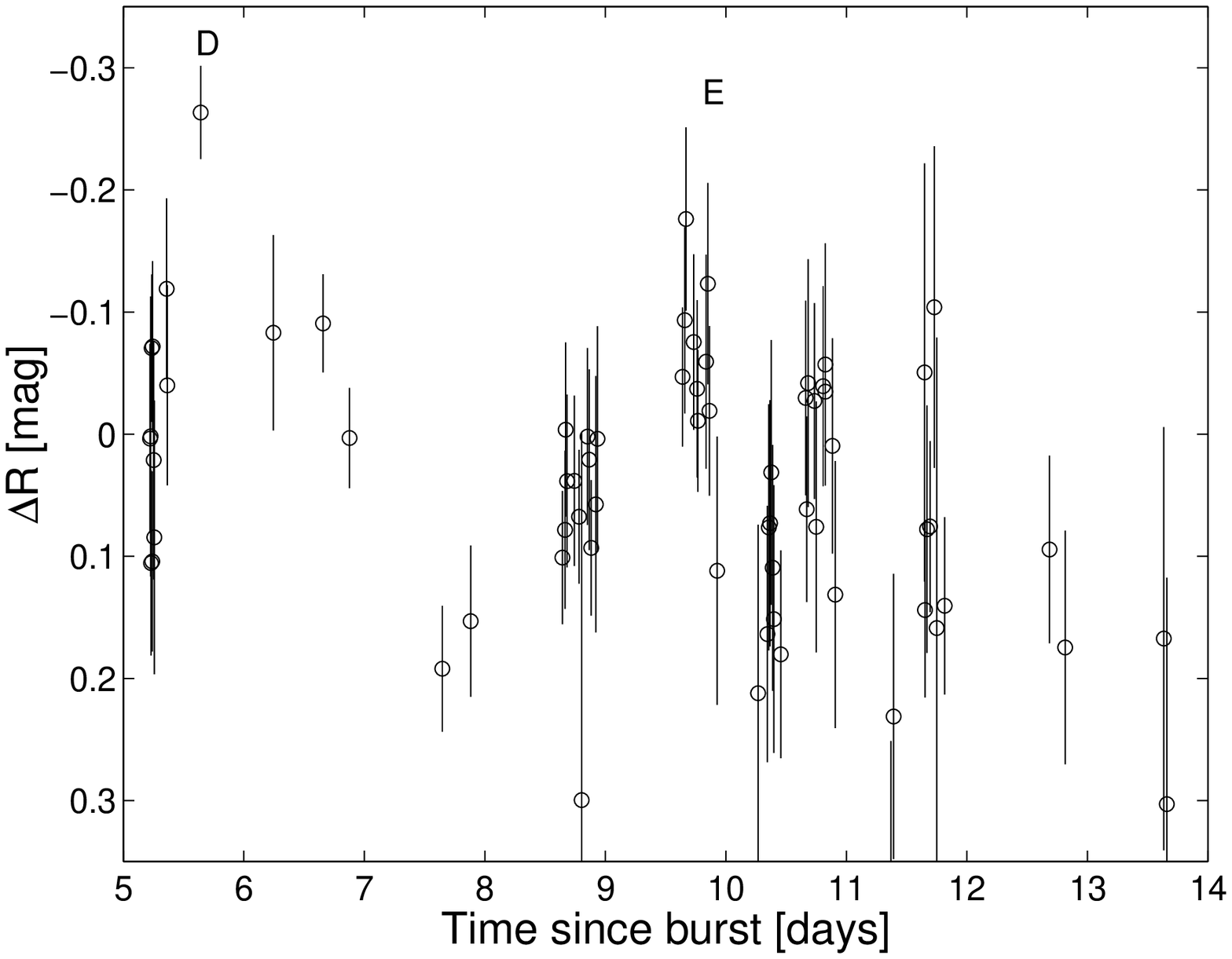}}
\caption{The variations light curve of the afterglow between day $5$ and day $14$.
}
\label{FigRes812}
\end{figure}

\begin{figure}
\centerline{\includegraphics[width=8.5cm]{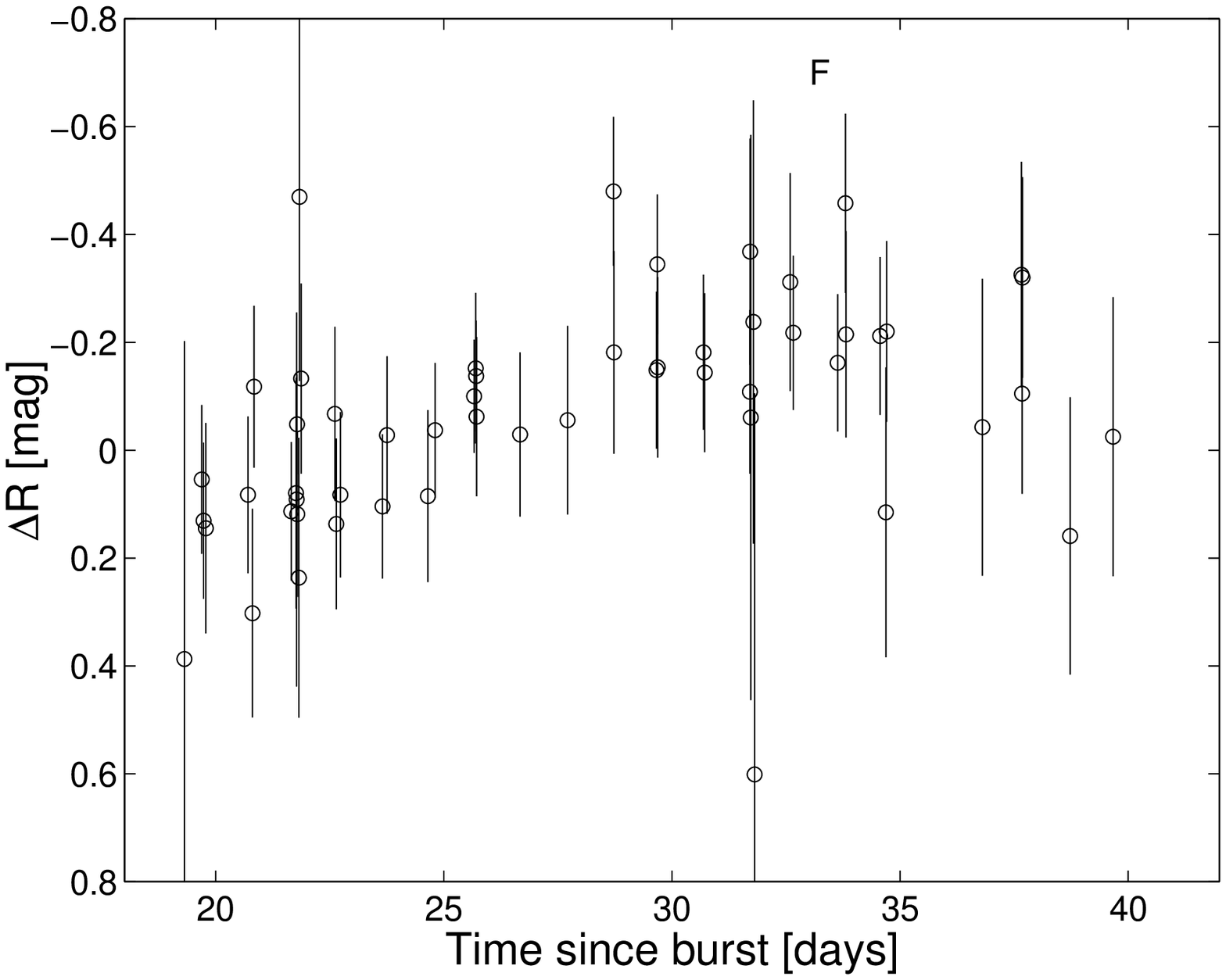}}
\caption{The variations light curve of the afterglow between day~$18$ day~$42$.}
\label{FigRes1842}
\end{figure}

Bumps $A$, $B$, and $C$ share a strikingly similar overall structure
and time-scale during $\sim0.6$~d about their maximum.
This similarity is demonstrated in Fig.~\ref{FigBumps} (panels {\it
  a}, {\it b}, and  {\it c}),where each of the three bumps is shown,
compared to a curve manifesting the common coarse structure of these
bumps.
To derive this ``standard profile'' of the bumps, we superposed bumps
$B$ and $C$ over bump $A$. 
Bumps $B$ and $C$ were shifted by $\Delta{T}=-1.02$~d,
$\Delta{m}=-0.008$~mag, and $\Delta{T}=-1.718$~d,
$\Delta{m}=0.055$~mag, respectively, to obtain a good match by eye.
The data were then smoothed using cubic splines.
Finally, we derived a profile curve using a high-degree polynomial
fit.\footnote{A tabulated version of the ``standard profile'' is
  available in our website,  http://wise-obs.tau.ac.il/GRB030329/}

The three events are consistent with this ``standard profile'',
comprising of a fast monotonic rise, and a slower, complex decline.
Panels {\it d}, {\it e}, and {\it f} in Fig.~\ref{FigBumps} show a
comparison between the  ``standard profile'' and bumps $\aleph$, $A'$,
and $D$, respectively.
The brightening rate of bump $\aleph$, as well as its time scale, are
similar to the standard profile.
However, its structure is significantly different.
In particular, in contrast to the concave early decline which $A$,
$B$, and $C$ seem to share, the declining branch of $\aleph$ has a
convex form.
Our fragmented data of bump $D$ are consistent with the standard
profile, but on a time scale that is longer by a factor of $\sim2-2.5$.
Finally, the rising branch of minor bump $A'$ seems to be similar to
the tip of standard profile, but its decline is slower, perhaps
because of the rising of bump $B$.

Another apparent phase of variability, between $8$~d and $13$~d,
features three consecutive low-amplitude ($\Delta{R}\approx0.1$~mag)
bumps, with a time-scale of $\sim1$~d (Fig.~\ref{FigRes812}).
However, the amplitudes of the later two bumps ($t\sim10.7$ and
$t\sim11.7$~d) are within the cross calibration systematic uncertainties, and
their maxima data all come from a single source (Mount Laguna
Observatory), and should be therefore treated with caution until confirmed by
other observations.
A broad bump ($F$) spanning $\sim20$~d occurred around $t\approx30$~d
(Fig.~\ref{FigRes1842}).
The shape of this feature is somewhat sensitive to the SN model used.
Nevertheless, the deviation from the power-law decline during this
period persists for any of the SN models which we have tested.

Strong variability is also detectable during the late decline, after
$t\sim 50$~d. In particular, a $\sim0.3$~mag rebrightening on a time scale of 2~d  
occurred around $t\approx52$~d (the ``jitter episode'' of Matheson et al. 2003; 
See Fig.~\ref{FigLCBreak}).
Our light curve features another $\sim0.2$~mag rebrightening on a time
scale of a few days, around $t\approx64$~d.
Because the observations tracing this variation were all obtained at the
same observatory (MDM 2.4m), it is unlikely that this  is due to some
reduction artifact.
To conclude, it appears that successful models of this well-observed
GRB should ultimately account for strong variations of the
optical emission, on time scales of hours to weeks, recurring over
tens of days after the burst.

\subsection{Color evolution}
\label{SecColor}

\begin{figure}
\centerline{\includegraphics[width=8.5cm]{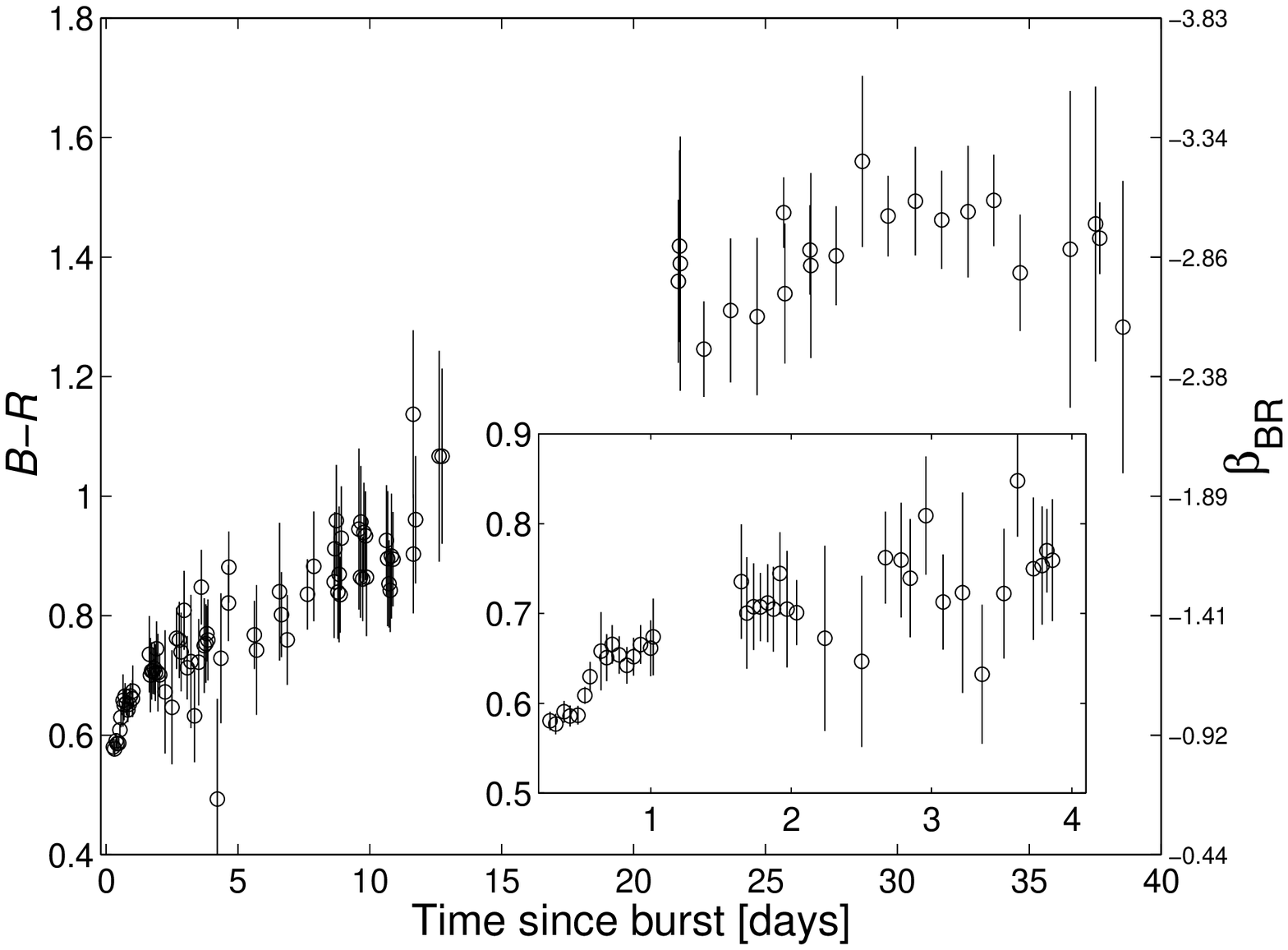}}
\caption{The binned $(B-R)$ color of the OT during the first $40$~days
  after the burst. The vertical axis on the right shows the
  corresponding spectral power-law, $\beta_{BR}$. The inset is a
  blow-up of the first four days.}
\label{ColorBR}
\end{figure}

Figure~\ref{ColorBR} shows the
$B-R$ color evolution of the light curve of GRB~030329.
The host and the SN light curve were not subtracted, in order to
reduce propagated errors and keep the results model-independent.
To derive the $B-R$~light curve we interpolated the better sampled $R$-band
light curve onto the times of the $B$-band light curve.
The interpolation uncertainty was calculated using the method of
 Ofek \& Maoz (2003).
Finally, we binned the $B-R$ color light curves
in $0.05$~d bins.

The color evolution of this event was studied in detail by
Matheson et al. (2003) and Bloom et al. (2003). Both groups
used subsets of the data presented here. Our analysis
yields consistent results with those previous works. In
particular, we measure almost constant colors between
days $2 - 5$, after which the emerging SN component
(with colors similar to SN 1998bw)
drove the color evolution of the optical emission. 
Furthermore, the earlier coverage
of our light curves enables us to measure the color evolution of the
OT during the first day. 

The $B-R$ light curve shows a small ($\sim0.1$~mag),
but significant, color variation during the first day
after the burst (Figure~\ref{ColorBR}, inset).
The color of the OT 
evolved from $B-R=0.58\pm0.01$~mag at $t=0.28$~d
(corresponding to $\beta_{BR}=-0.88\pm0.02$)
to $B-R=0.66\pm0.02$~mag at $t=0.83$~d
(corresponding to $\beta_{BR}=-1.07\pm0.05$),
where $\beta_{BR}$ is the spectral energy slope
defined by, $f_{\nu} \propto \nu^{\beta}$ ($f_{\nu}$ is the specific
flux). 
This early evolution is unlikely to be related to the SN component. 
To date, color evolution has been observed in the light curves
of GRB~021004, starting $\sim1.5$~d after the burst
(Bersier et al. 2003; Mirabal et al. 2003),
and possibly also in GRB~000301C (Rhoads \& Fruchter 2001).
An interpretation of the
color change of GRB030329 in the context of the relativistic 
synchrotron model, as a manifestation of the cooling break
frequency going through the optical bands (e.g., Galama 
et al. 2003) is discussed below (see \S\ref{secbreaks}).

\section{Discussion}

\subsection{Early breaks}
\label{secbreaks}

The relative proximity of GRB~030329 has prompted special
attention by the astronomical community, and several works
presenting analysis of various data sets have been published so far.
Our superior compilation of optical data, as well as the privilege
afforded by the availability of these earlier works, allow us to
inspect some previous suggestions in light of the newly available
data.

Early analysis of preliminary optical data revealed that the slope of
the optical decline became steeper around day $0.5 - 0.6$
(e.g. Garnavich, Stanek, \& Berlind 2003; Burenin et al. 2003; Price
et al. 2003b).
The change in the power-law decline index  from $\sim1$ to $\sim2$,
seen both in early optical data and  in sparse X-ray observations
reported by Tiengo et al. (2003), combined with the achromatic nature
of the break (Burenin et al. 2003),  seemed to support an
interpretation of this steepening as a  ``jet break'' - the
manifestation of a conical geometry in the relativistic emitting
material (e.g., Rhoads 1997). 

\begin{figure}
\centerline{\includegraphics[width=8.5cm]{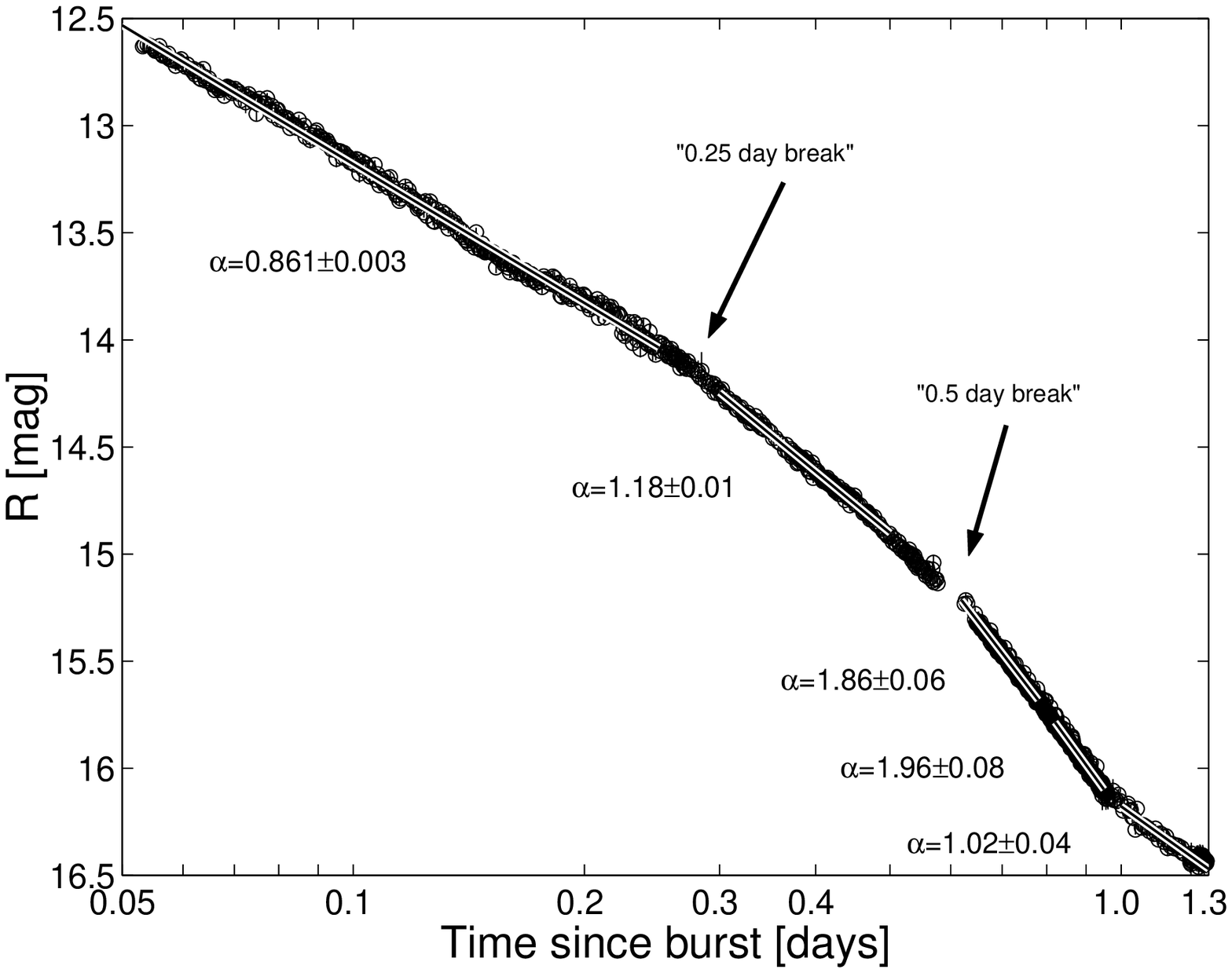}}
\caption{A detailed view of the early light curve of the OT
associated with GRB~030329. 
The light curve distinctly breaks around $0.25$ and $0.5$ days.
The data obtained prior to $0.25$ days is
not well described by a smooth power law, and shows
significant wiggles.
Note that our data rule out further steepening 
of the light curve at day $0.8$, with the slope measured
during $0.82-0.95$ days consistent with the
slope measured from earlier data ($0.6-0.78$ days)
and certainly below the values $\alpha\sim2.2$
reported elsewhere (see text). Around day $1$ the decline
in the optical flux begins to slow, as the
rising branch of bump $A$ emerges.}
\label{EarlyLc}
\end{figure}

Figures~\ref{FigLCBreak} and~\ref{EarlyLc} show that this simple 
interpretation does not fit the well sampled light curves now available. 
A model postulating a steep ($\alpha\sim2$) optical decline starting
at day $\sim0.5$ would severely under-predict the optical observations
from day 1.5 onward. Sustaining a model involving a ``jet break'' at
day 0.5 requires an additional source of optical emission emerging
at day $\sim1$. Such a model was indeed suggested by Granot et
al. (2003) and Berger et al. (2003) and will be discussed in the next
section.

As can be seen in Figures~\ref{FigRes07} and \ref{EarlyLc}, 
as well as in Uemura et al. (2003) and Sato et al. (2003), the early optical 
data (i.e., before the suggested break at day $\sim0.5$) are not
consistent with a single power-law decline, as predicted before
a jet break.
In particular, the data presented by Uemura et al. (2003),
Torii et al. (2003), and Sato et al. (2003) require
at least one additional break to occur around day $\sim0.25$.
Our compilation clearly elucidates this (Fig.~\ref{EarlyLc}). 
Several authors (Torii et al. 2003; Sato et al. 2003) 
suggested that this additional break (with a change in the temporal slope of
$\Delta\alpha\sim0.3$) may represent the so-called ``cooling break'',
predicted in the context of relativistic fireball models
(Sari, Piran \& Narayan 1998) to occur as
the cooling-break frequency, $\nu_c$, passes through the optical band
(with $\Delta\alpha=0.25$). As seen in Fig.~\ref{EarlyLc},
the combination of an early cooling break at day 0.25
and a jet break at day 0.5 allows a fair representation
of the {\it early} optical data. Thus, successful models
of this event should account for at least two early 
breaks. However, it should be stressed that the well
sampled light curves available before $0.25$ day are
not consistent with a smooth power-law decline, and 
show significant wiggles and bumps (we shall further discuss 
this point below).

Previous studies suggesting an early cooling break relied on
the analysis of single-band data, and could therefore
only detect the predicted shift in the temporal
decay slope $\alpha$, as discussed above. Our compilation also
allows us to inspect the color evolution of the OT
around the time of the suggested cooling break.
As reported in \S\ref{SecColor}, we detect a shift in the optical 
spectral index of $\Delta\beta=0.19\pm0.05$ between $0.28$ (our earliest
available color data) and $0.83$~d after the burst.

Relativistic synchrotron models predict that a cooling break sweeping
the optical band would manifest itself by a change in the color index
of $\Delta\beta=0.5$.
If the observed color change is indeed due to the cooling break, then
considering that $\sim40$\% of the expected color change occured
between $0.28$ to $0.83$~days, and noting that both theory (Sari et
al. 1998) and observations (Galama et al. 2003) show that the cooling
break evolves in time with a power law, we estimate that the cooling
break crossed the $V$ band at day~$\sim0.25$.
We note that the observed steepening of the slope passed through the
optical band from high frequencies to low ones, and hence demotes
models predicting an opposite trend.

Another way to probe the cooling break frequency value 
during later times is through the optical 
to X-ray spectral index, which is predicted to be constant
after the cooling break moves below the optical bands.
We measured this in four epochs of XTE and XMM-Newton observations
given by Tiengo et al. (2003; $t=0.222$, $1.24$, $37.24$, and
$60.85$~d; the first epoch is the weighted mean of four
measurements), by interpolating the $R$-band afterglow light curves onto the
four epochs of the X-ray observations.

The value of slope at the first epoch 
$-0.93\pm0.02$ (at $t=0.222$), before or
during the proposed cooling transition, markedly
differs from the values obtained at later epochs, after the break,
$-1.04\pm0.05$ and $-1.06\pm0.13$ at $37.24$ and $60.85$~days after
the burst, respectively.
The value measured $1.24$ days after the burst, $-0.95\pm0.04$, 
is consistent with both early and later values.  
A coherent picture therefore emerges from the optical observations
and X-ray data in which the color evolution we
detected is possibly the result of the cooling break passage through the
optical band around $0.25$ days after the burst. 
Naturally, other explanations for the observed early color evolution
are possible. 

Finally, we consider the report by Smith et al. (2003) of 
a further steepening of the optical light curve 
(reaching $\alpha\sim2.2$) around 0.8 day -- distinct from, and
occurring after the proposed $0.5$ day break.
This could have confirmed the prediction
by Granot et al. (2003) that the final slope after
a jet break around $0.5$ days should be steeper than
the value reported ($\alpha=1.9$), as expected from
the temporal slope change due to a jet break,
$\Delta\alpha\sim1$, relative to the post-cooling break 
slope $\alpha\sim1.2$. However, our superior data, including
two mutually consistent data sets from the MDM and Mount
Laguna Observatory, obtained through standard filters, are
not consistent with the unfiltered data reported by Smith et al. (2003).
Fig.~\ref{EarlyLc} shows no segment of the early light curve (including
the period between $0.8-1$ days) with
a power-law slope steeper than $\alpha=2$. Thus, the
theoretically predicted slope of $\alpha=2.2$ was either
not reached, or washed out by the emerging bump $A$.    
    
To conclude, our results support the interpretation of the
break around $0.25$ days, as a cooling break, within the context of
relativistic synchrotron models.
The interpretation of later steepening of the light curve (around 
$0.5$ day) as associated with a geometric (``jet'')
break, can be sustained only if an additional emission source,
dominating the optical flux from day $1.5$ onward, is invoked.

It should also be noted that Dado, Dar, \& De R{\'u}jula (2003) also
put forward a theoretical prediction for the optical light curves
of this event, based on their ``cannon ball'' model.
Lacking access to the predicted
curves, we are unable to conduct a detailed comparison with our
data. However, such a comparison is certainly warranted, and could be easily
conducted using our publicly available data.

\subsection{The two-jet model revisited}
\label{Berger}

Berger et al. (2003) analyzed radio observations of GRB~030329,
and found that the radio data, as well as mm-band observations reported
by Sheth et al. (2003), could be well 
described by a relativistic jet model. However, the jet parameters
indicated a ``wide'' jet, exhibiting the characteristic jet break
about 10 days after the burst. In view of previous
analysis of early-optical and X-ray data which seemed to require a ``narrow'' 
jet, breaking around day $0.5$, Berger et al. (2003) proposed
a composite two-jet model, combining a narrow ultra-relativistic 
component responsible for the $\gamma$-rays and early ($t\ltorder1.5$ days)
optical and X-ray afterglow, with a wide, mildly relativistic
component responsible for the radio and optical afterglow beyond 1.5
days.
This model provides a good fit to the radio and mm data as well as to
the preliminary optical data used by these authors.
The model also describes well the early X-ray observations, but
under-estimates late ones.
The wide jet component provides the additional source of emission
required to account for the optical observations after day $1.5$, as
discussed above.

The energy derived from the Frail relation (Frail et al 2001; Bloom et al. 2003),
assuming only a narrow-jet component and the updated parameters used by
Bloom et al. (2003), falls $7\sigma$ below the 
geometrically corrected mean gamma-ray energy of Bloom et al. (2003).
The contribution of the wide-jet component introduced by Berger et al. (2003)
brings the total energy to within $1\sigma$ of the mean value of Bloom et al. (2003).

\begin{figure}
\centerline{\includegraphics[width=8.5cm]{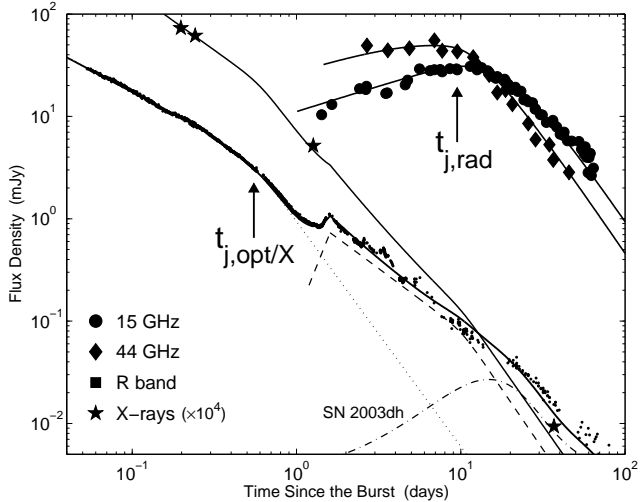}}
\caption{A comparison of the updated two-jet model with multi-band
data. The model and notation are similar to those presented by Berger
et al. (2003), except for several minor modifications (see text). 
The various symbols denote data points, as labeled, and the solid
curves are the model predictions in each respective band. The
model $R$-band light curve describes the general trends in the 
observed light curve well, but some discrepancies remain -- the model becomes
too bright around 10 days after the burst, and later on becomes
two faint.}
\label{bergerfig}
\end{figure}

With our improved optical light curves at hand, we revisit the Berger
two-jet model. Figure~\ref{bergerfig} shows a comparison between an
updated two-jet model with radio data from Berger et al. (2003) and X-ray 
data from Tiengo et al. (2003), as well as our R-band light curve. The
model is essentially the same as the one described in Berger et al. (2003),
except for the following modifications. 
Firstly, a cooling break component at $0.25$~day (see~\S\ref{secbreaks})
was incorporated into the model.
Additionally, the temporal emergence of the second jet at
$t\sim1.5$~day was set to $t^4$ (instead of the $t^2$ law used in
Berger et at. 2003) to account for the abrupt rise which our nearly
continuous data show (in fact, an even steeper emergence is probably
required).
Finally, the 1998bw-based SN model used by Berger at al. (2003), which
we showed to be too bright, was replaced by our best SN
model as described in \S\ref{SecSNLC}.

\begin{figure}
\centerline{\includegraphics[width=8.5cm]{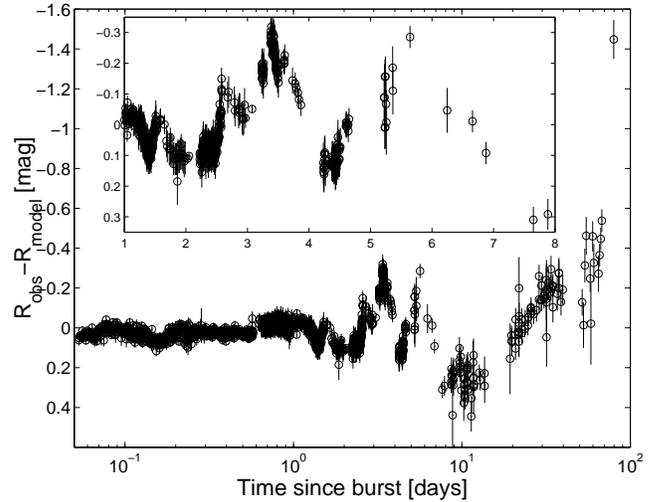}}
\caption{The residual light curve obtained by subtracting the modified
two-jet model from the observed $R$-band light curve. Note that at
$t\le1$~day the model describes the general light curve trends very
well, but significant wiggles around the smooth trend remain. 
A phase of strong variations can be seen between $1-8$ days,
followed by broader undulation around the predicted peak of
SN 2003dh, that may be associated with inadequacies in the 
SN model. During later times ($\gtorder20$ days), the data seem to 
require yet another source of optical emission, which is also
strongly variable.
The inset is a magnified view of the strong variations between, day
$1$ and day $8$}
\label{bergerresall}
\end{figure}

\begin{figure}
\centerline{\includegraphics[width=8.5cm]{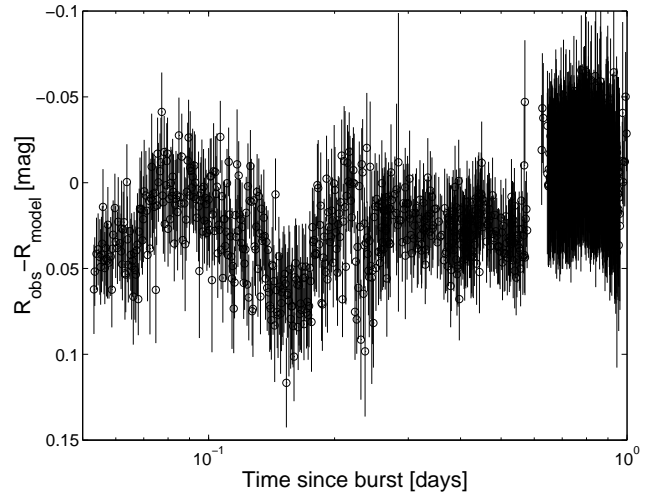}}
\caption{Same as Fig.~\ref{bergerresall}, but zooming in on the
period prior to the emergence of bump $A$. The residual undulations
after subtraction of the Berger et al. (2003) two-jet model 
light curve are clearly evident.}
\label{bergerresearly}
\end{figure}

The modified two-jet model fairly describes the general trends in our
well sampled optical light curves.
However, some discrepancies still remain. 
Figure~\ref{bergerresall} shows the residuals obtained by subtracting the
updated two-jet model from the observed $R$-band light curve.  
At early times ($\ltorder1.5$~day), the model fits the trends in the light
curve.
Nevertheless, significant undulations (with peak-to-peak $\gtorder0.1$
mag), are detected about the smooth model (Fig.~\ref{bergerresearly}).  
Thus, this, or any other model based on broken power-law segments, should
ultimately be supplemented with a mechanism explaining the bumpy
nature of the optical emission, starting at the very earliest times
(e.g., even prior to the cooling break). 
Following, a period of strong variations ensues, lasting till day eight. 
Comparing the inset in Figure~\ref{bergerresall} with Figure~\ref{FigRes07}, 
we see that the two-jet model eliminates bump $\aleph$ (which under this
interpretation is the combined effects of the cooling and jet
breaks) while bump $A$, associated with the emergence of the second jet,
is diminished. 
However, later structure, including bumps $B$, $C$, and $D$, remains. 
Moreover, both the structure of these bumps and their peak-to-peak
amplitude, are essentially unaltered.
Thus, the modified two-jet model does not eliminate the need for an
additional, strongly variable component of optical emission.  

After day eight, a smooth undulation, lasting till day $\sim20$, is seen in
Figure~\ref{bergerresall}. Since this is the period in which the
optical flux is 
dominated by the emission from SN 2003dh, the discrepancy seen may
indicate that
our model light curve of SN 2003dh is at fault. This undulation can be
almost completely removed by making our model for SN 2003dh fainter by
0.3 mag --  at the lower limit of the range we consider plausible (see
\S\ref{SecSNLC}). 

Finally, the data clearly require an additional source of optical flux in order
to explain the late-time ($\gtorder 20$ days) light curve. This extra
component (in addition 
to the narrow and wide jets, and SN 2003dh) should {\it rise} from day $20$ 
and through day $79$, where our data end, and is also required to be strongly
variable in order to explain the late time light curve
``jitters''. This extra flux
cannot be attributed to the SN without making its light curve very different
from that of SN 1998bw, and highly variable. Further tests for the relative 
contribution of SN~2003dh to the late-time optical flux can be obtained 
from late-time spectroscopy of this event. Interestingly, as can be
seen in the lower-right corner of Figure~\ref{bergerfig}, the modified 
two-jet model also under-predicts the late-time X-ray points (at $t=37.24$,
$60.85$~day), perhaps indicating
a need for an extra source of late-time flux in this band also.     

To conclude, as already shown by Berger et al. (2003), the two-jet model
provides a fair, self consistent description of observational data from the 
radio to the X-ray. However, in view of the large volume and complex
nature of the multi-band data obtained for this event, further investigation
of this model, perhaps combined with a prescription for the additional
variations component required, is warranted. 

\subsection{Can a single broken power-law provide a good fit
to the data ?}
\label{onebreak}
We demonstrated above that if we assume a power-law decline -- the
simplest OT model commonly used, the data require different slopes for
the early- and late-time decline slopes (\S~\ref{SecAGLC}).
The values of the decline indices, $\alpha_1\approx1.1$ and
$\alpha_2\approx2$  are robustly constrained by the data. 
The transition between these decline slopes requires a break in the
optical light curve $\sim3$ to $\sim8$ days after the burst.
Unfortunately, the strong undulations superposed on the smooth decline
trend throughout this period prevent us from determining the
accurate timing of the power-law break.

It is intriguing to test whether this simple model, which naturally
emerges from the analysis of the optical data,
can consistently provide a reasonable fit to the entire 
multi-band data set gathered for this burst. In this context, we
note the following major points.   
In the optical regime, this simple model for the ``smooth'' evolution 
of the OT requires additional emission component/components
to account for the strong flux variability detected from hours to
weeks after the burst. 
The ubiquity of these undulations suggests that such a component
cannot be avoided by more complex models which have thus far been
proposed.
Particularly, as  shown  (\S\ref{Berger}), the two-jet model advocated
by Berger et al. (2003) accounts for some of the more 
prominent features in the optical light curves of this event ($\aleph$
and $A$ in Fig.~\ref{FigRes07}), but does not account for other, equally
prominent ones.
Thus, without concrete and self-consistent  models for the mechanism
of the light curve variability, the amplitude of the undulations about
the simple broken power-law model for this burst does not seem to
argue against the validity of the single break model.

The X-ray coverage of this burst is regrettably sparse.
Comparing the available data reported by Tiengo et al. (2003)
with our optical light curves, and assuming that the X-ray and
optical flux are correlated, as found by Fox et
al. (2003) for GRB~021004, we expect the optical to X-ray 
slope index to remain approximately constant. As we have
shown in \S\ref{secbreaks}, this is indeed the case, especially
if the early X-ray point, obtained before the proposed
cooling transition, is discarded. We note, however, that
when over-plotting the X-ray data over the $R$-band
light curve, and scaling the X-ray points to match the early
light curve, the two latest XMM points fall below their
expected position, by a factor of $6-10$.
The physical significance of this discrepancy is not clear.
We note, however, that a discrepancy of similar
magnitude (but of opposite direction) is found also when comparing the
modified two-jet model (\S\ref{Berger}) with the X-ray data
(Fig.~\ref{bergerfig}, bottom right corner).
If any of these models is correct, this may suggest a weak
evolution in the optical to X-ray spectral slope.

Finally, it appears that the greatest challenge for the simple
one-break model is accounting for the results of the radio
and mm observations.
As elaborated by Berger et al. (2003), these data require the
existence of a mildly-relativistic jet, which is expected to
demonstrate a break in the optical regime around day $10$ after the
burst.
Thus, the single break between $\sim3$ to $\sim8$~d after the burst
implied by the simple model seems to be somewhat in odds with the
constraints posed by the radio and mm data.
However, no confidence intervals have been determined for the timing
of the break, both in the optical and in radio.
In the optical band, the measurement is hindered by strong, multiple
bumps and wiggles superposed on the  smooth light curves.
The apparent conflict in the determination of the break time may
perhaps be negotiated by future analysis, which would also model
the variations in the  optical band (e.g., by fitting a physically
motivated model explaining the observed optical light curves). 
We thus conclude that the data at hand cannot rule out the
empirically motivated singly-broken power-law model for the emission
associated with GRB~030329.
Assuming this model with $t_{break}\approx8$~day,  $E_{iso}(\gamma)$
as found in \S\ref{Berger}, and an interstellar matter density
$n=1.8$~cm$^{-3}$ (Berger et al. 2003), the estimated total energy of
the event,  derived using the Frail relation (Frail et al. 2001; Bloom
et al. 2003), is $4.4\times10^{50}$~erg.
This value is within $1.5\sigma$ of the updated geometrically
corrected mean gamma-ray energy value of Bloom et al. (2003).  

\subsection{Light Curve Variations}
\label{Secdiscvar}

As shown above, the optical light curves of the OT associated with 
GRB~030329 exhibit strong undulations superposed on the overall 
smooth trends, detectable shortly (hours) after the burst, and still 
apparent many tens of days later.
The variations seen during the period which is best-sampled by our
light curves have a typical time scale of $\sim1$ day, and show a
similar asymmetric structure, with a rising branch about twice as
steep as the declining one.
Both the characteristic amplitude ($\sim50\%$), and the structure of
these variations are only weakly influenced by our assumptions about
the underlying smooth behavior.
In particular, similar results are obtained when we assume
either the simple, empirically motivated, singly broken power-law model,
or the more complex two-jet model of
Berger et al. (2003; Figures~\ref{FigRes07} and \ref{bergerresall},
see inset).
 
Similar undulations have been previously observed in the light curves
of OTs associated with GRBs.
A short-time variation was detected in the light curves of
GRB~000301C (e.g., Masetti et al. 2000; Rhoads \& Fruchter 2001; Berger et
al. 2000; Sagar et al. 2000).
The short time scale and achromatic nature of the variation led
Garnavich, Loeb, \& Stanek (2000) to suggest this bump was caused by
microlensing of the OT by a star in a foreground galaxy, while a more
mundane origin for the bump -- nonuniform ambient density -- was
proposed by Berger et al. (2000). 
The time scale and structure (fast rise, slow decline) of the bump 
detected in GRB 000301C, are similar to those seen in our light curves
of GRB~030329. 
This, combined with the different structure seen in optical and IR
light curves of the OT of GRB~000301C (Sagar et al. 2000), which also
leads to a poor fit of the multi-band microlensing model to the
observed $R$-band light curve (Garnavich et al. 2000) suggests that
the bump seen in the light curve of GRB~000301C is more likely to be
intrinsic to the source, than the result of the rare cosmic alignment
required for microlensing.
More recently, numerous bumps were detected in the light curve of the
OT associated  with GRB~021004 (e.g., Bersier et al. 2003; Mirabal et
al. 2003; Fox et al. 2003).
Here too, the temporal structure and time scale of the bumps is
similar to those seen in our light curves of GRB 030329.

A seemingly exceptional case to note is that of GRB~970508, the second
burst for which an OT was identified.
The optical transient of this burst (e.g., Galama et al. 1998)
showed a major rebrightening around one day after the burst. 
Following a short rise lasting less than a day, the optical emission
underwent a smooth, power-law decline over $\sim100$ days. 
The relatively sparse sampling of the light curve of this burst after
day $10$ does not allow us to determine the exact nature of the light
curve decline during late phases (e.g., search for late ``jitter''
periods), but the overall smooth structure and very large amplitude of
this rebrightening appear to be quite different from those of the OT
associated with GRB~030329 (or from any other OT so far detected).

The common explanations for the short time-scale variations
invoke complex density  structures around the burst (e.g. Wang \& Loeb
2000; Berger et al. 2000; Lazzati et al. 2002), inhomogeneous energy
disposition within relativistic conical blast waves (the ``patchy
shell'' model; Kumar \& Piran 2000a) and continued injection of energy
by the central engine (refreshed shocks; e.g., Rees \& M\'esz\'aros 1998;
Kumar \& Piran 2000b).
Nakar, Piran \& Granot (2003) performed a detailed
theoretical investigation of GRB~021004, and showed that variants
of all these models can fit the data, although they preferred a version
of the ``Patchy Shell'' model. 

Granot et al. (2003) next studied the variability of GRB~030329, using
a preliminary compilation of early optical data. Interpreting our
detailed light curves in view of their analysis, we note the
following points.
The significant variations observed in the light curve before the
proposed cooling break at day $0.25$ probably argue against variable
density models, in agreement with the conclusions of these authors.
Interpreting the $0.5$ day steepening in the light curve
as a jet break, Granot et al. (2003) ruled out the patchy shell model,
since it cannot produce strong variations
after the entire jet is visible (i.e., after the jet break).
Since strong variations are observed also well after day $10$
(e.g., the ``jitter episode''), the patchy shell model cannot explain
the late variations observed in the light curve of GRB 030329.
However, if the jet-break occurred as late as day eight (e.g., as in
the single power-law model), the patchy shall model may explain the
strong early variations.

The refreshed shocks scenario (Kumar \& Piran 2002b),
predicts that if the shocks occurred before the jet spreading, the time
scale of the variations would be $\Delta{t}\sim t$.
Granot et al. (2003) showed, however, that if the refreshed shocks occurred
after the jet spreading, the time scale of the variations would be
$\Delta{t}\sim t^{1/4}$. 
Considering the roughly constant time scale of the variations,
and assuming a jet-break at day $0.5$,
Granot et al. (2003) favored the refreshed shocks model for
the variations in the light curves of this burst.
However, if a jet-break occurred after day $\sim$five, the almost
constant time scale of the variations would be in contrast with the
refreshed shocks model advocated by Granot et al. (2003).
Inverting this argument, if refreshed shocks are shown to be the
likely mechanism causing the observed variations, a model with a
single jet break around day five becomes unphysical.

The similar structure of the undulations seen in the light curve of
GRB~021004 (Nakar et al. 2003), GRB~000301C 
(Panaitescu et al. 2001) and GRB030329,
may hint on a ``standard'' variability mechanism in OTs.
Naturally, many more cases should be studied in order to confirm this
suggestion.

\section{Summary}
\label{Summary}
We observed the optical afterglow of the nearby GRB 030329
from five observatories across the globe.
We carefully cross-calibrated the observations, and augmented them
with published data.
The final compilation of $BVRI$ light curves is unprecedented in its
temporal sampling, and reveals complex structure.

Decomposing the light curve into host galaxy, SN, and afterglow
components, we showed that SN 2003dh, associated with GRB~030329,
could not have had a light curve identical to that of SN~1998bw.
Instead, the evolution of SN 2003dh is better described by making the
light curve of SN 1998bw fainter by 0.3 mag, and with a time scale
that is 0.8 times shorter.
We subtracted this SN model from the light curve of GRB~030329, and
found that the residual light curve is well described by a double
power-law, with a break point in the range of $\sim3$ to
$\sim8$~days.
The power-law slope of the light curve changed from
$\alpha_{1}\approx1.1$ to $\alpha_{2}\approx2.0$.
These results are very weakly dependent on the SN model used.

The SN, host-galaxy, and power-law subtracted light curves of
GRB~030329 show strong variations with time scales ranging from
$\sim0.5$~hr to $\sim10$~d.
The early variations ($\ltorder8$ days) which are well covered by our
observations, are typically asymmetric with their ascending branch
about twice as fast as the descending branch.
Their typical, peak-to-peak time scales are $12-24$~hr.
Three of the bumps ($A$, $B$, and $C$) are of a similar structure
during $\sim0.6$~days about their maximum.
Later variations are harder to characterize, because of the lower
frequency of our sampling.
Nonetheless, they seem to have a longer time scale.
Periods of strong variability are still evident in the light curve
tens of days after the burst.

We showed that the OT color changed during the first day after the
burst.
In the context of relativistic synchrotron models, this supports the
suggestions, based on single-band light curves, that the ``cooling
break'' frequency passed through the optical bands around $0.25$
days.

We discussed previous analysis of this event and found that the 
simple model involving a ``jet break'' occurring around day 0.5,
proposed by several authors, is unable to account for the detailed 
optical data available for this event. At least two additional
emission sources
are required in order to sustain this interpretation:
one to account for the
flux observed from day one onward, as done by Berger et al. (2003),
and another mechanism must be invoked to account for the ubiquitous
undulations detected in the light curves throughout the observed
period, i.e., both before and after the jet break.

Examining an updated version of the Berger et al. (2003) model, we
find that it provides a fair description of available multi-band data.
However, several discrepancies still need to be accounted for,
probably requiring a self consistent model for the mechanism
producing the ubiquitous bumps and wiggles in the light curves of
this event. A similar effort is required in order to test 
a simpler, empirically-motivated singly-broken 
power-law model we present. 

A comprehensive effort to
model the large volume of data collected for this burst, 
explicitly accounting for the variability on all time scales,
appears like the next challenge in making this
unique event a key to understanding the GRB phenomenon. In this vein,
in order to ensure the maximal usefulness of our observations to the community,
we make all the data available through our web site:
$http://wise-obs.tau.ac.il/GRB030329/$.

\acknowledgments
We thank,
M. Uemura and R.A. Burenin
for sending us a digital version of their
light curves, and the authors of the 
paper by Matheson et al. (2003) for the early release
of their data to the community. 
We are grateful to A. Fruchter for sending us the host
galaxy magnitude prior to publication.
We acknowledge fruitful discussions with E. Waxman,  D. Frail, and
A. Levinson. 
YML and EOO are grateful to the Dan-David prize foundation
for financial support. AG acknowledges a Colton Fellowship.
YML, EOO, AG, and DP $\times 2$ were supported in part by grants from
the Israel Science Foundation.
We thank A. Henden and the USNO staff for providing us,
and the GRB community in general, with field photometric calibration.
We thank Eve Armstrong for observations at MDM, as well as
the observers and staff of all the observatories who
participated in this campaign, without which this work would
not have been possible.

\end{document}